\documentclass[
aip,
pof,
amsmath,amssymb,
preprint,%
]{revtex4-2}

\usepackage{latexsym}
\usepackage{hyperref}
\usepackage{fancyhdr}

\usepackage{graphicx}% Include figure files
\usepackage{soul,color}
\usepackage[dvipsnames]{xcolor}
\usepackage{dcolumn}% Align table columns on decimal point
\usepackage{bm}% bold math
\begin{document}

	%\preprint{AIP/123-QED}
	
	\title{A new compact active turbulence generator for premixed combustion: Non-reacting flow characteristics}   % Force line breaks with \\

	\author{Sajjad Mohammadnejad, Leslie Saca, and Sina Kheirkhah}
	\email[Author to whom correspondence should be addressed:~]{sina.kheirkhah@ubc.ca}	
	\affiliation{School of Engineering, The University of British Columbia, Kelowna, British Columbia, Canada, V1V 1V7}

	%\date{\today}% It is always \today, today, but any date may be explicitly specified

\begin{abstract}
A new compact active turbulence generator is developed, tested, and characterized, which extends the capabilities of such generators used in turbulent premixed combustion research. The generator is composed of two blades that resemble the shape of two bow-ties. Hot-wire anemometry and high-speed imaging are performed to characterize the non-reacting flow produced by the generator and the blades dynamics, respectively. Two mean bulk flow velocities of 5.0 and 7.0~m/s are examined. For comparison purposes, in addition to the developed generator, tests are also performed for a free jet as well as one and two perforated plates. The results show that the centerline root mean square velocity fluctuations can become about 1.8~m/s, which is several folds larger than that in the literature for reacting flows generated by active grids. For the newly developed device, the power-law decay of the one-dimensional kinetic energy is -1.0 and -1.3 for the mean bulk flow velocities of 5.0 and 7.0 m/s, respectively. The normalized energy dissipation rate is smallest for the newly developed device, while the energy dissipation rate is the largest. The spectral analysis of the velocity data does not show dominant frequencies equal to the blades rotation frequencies; and, the one dimensional kinetic energy and dissipation spectra follow -5/3 and 1/3 power-law relations. It is shown that, the small eddies produced by the newly developed device dissipate the turbulent kinetic energy faster than those corresponding to the rest of the tested turbulence generators.
\end{abstract}
	
\maketitle

\section{Introduction}\label{intro}

The \textit{modus operandi} of several existing land-based power generation gas turbine and possibly future aircraft engines is turbulent premixed combustion. Several review papers~\cite{lipatnikov2002turbulent,driscoll2008turbulent,driscoll2020premixed,steinberg2021structure,kheirkhah2021revisit} have been published during the past decades discussing the characteristics and dynamics of turbulent premixed flames. In majority of past investigations (including those~\cite{kheirkhah2016experimental,mohammadnejad2019internal,mohammadnejad2020thick} of the authors), passive turbulence generators, such as perforated plates, are used to produce turbulent premixed flames. For such turbulence generation equipment, the Root Mean Square (RMS) velocity fluctuations, $u'$, positively relates to the mean bulk flow velocity ($U$). Thus, maximum achievable $u'$ is usually limited by the maximum achievable $U$ in the experimental settings. The present study is motivated by the need to address this limitation via an active turbulence generator. A brief review of the literature concerning both passive and active turbulence generators are discussed in the following.

Passive turbulence generators used in combustion research can be broadly categorized as mesh screens/perforated plate(s), fans, slots, and combination of slots and impinging jets. Mesh screens~\cite{ballal1979influence,yoshida1980experimental,cheng1991influence,kalt2002experimental,chaudhuri2008blowoff}, one perforated plate~\cite{buschmann1996measurement,kobayashi1997turbulence,chen2003turbulence,smallwood1995characterization,lawn2006scaling,halter2009analysis,littlejohn2010laboratory,kheirkhah2013turbulent,chowdhury2017experimental,meehan2018flow}, and combination of perforated plates~\cite{fragner2015multi,mohammadnejad2021contributions,kheirkhah2015consumption} have been used in several past investigations. Combination of perforated plates refers to either a series of plates positioned along the flow direction~\cite{fragner2015multi,mohammadnejad2021contributions} or two perforated plates attached back-to-back~\cite{kheirkhah2015consumption}. Review of the literature shows that the attached perforated plates can produce relatively large values of $u'/U$ (about 20\% at 1.5 times burner diameter downstream of the turbulence generator). 

Fans are used in vessels for studying the effect of turbulence on spherically expanding premixed flames~\cite{groff1987experimental,kwon1992flame,koroll1993burning,shy2008effects,galmiche2014turbulence,morsy2021numerical}. For the turbulence produced by the fans, the RMS velocity fluctuations is positively related to the fans rotation speed. Such configuration is useful for controlling the turbulent flow characteristics, and the spherical flames have been used during the past decades for burning velocity estimation and model development. Although these are of significant importance, a mean bulk flow is absent in spherical flames, which makes comparisons with other flame configurations (which are relevant to e.g. gas turbine engines) challenging.

The slotted plates are used in the studies of Marshall \textit{et al.}~\cite{marshall2011development}, Videto \textit{et al.}~\cite{videto1991turbulent}, and Coppola \textit{et al.}~\cite{coppola2010experimental,coppola2009highly}. The combination of slotted plates and impinging jets has been used in the studies of Driscoll, Skiba, and Wabel~\cite{skiba2017structure,skiba2018premixed,wabel2017turbulent,wabel2017measurements,wabel2017experimental}. Comparison of the results presented in the literature shows that, amongst the tested passive turbulence generators, the combination of impinging jets and slotted plates leads to the largest values of RMS velocity fluctuations. For example, Wabel~\cite{wabel2017experimental} achieved $u'$ and $u'/U$ as large as 38~m/s and 43\%, respectively, using combination of slotted plates and impinging jets.

Despite active turbulence generators are used in many non-reacting flow studies~\cite{hideharu1991realization,mydlarski1998passive,larssen2011generation,bodenschatz2014variable,thormann2014decay,hearst2015effect,neuhaus2020generation,jooss2021spatialb}, such generators are rarely used in reacting flows. To the authors'  best knowledge, only Mulla \textit{et al.}~\cite{mulla2019interaction} has investigated the application of active turbulence generators for turbulent premixed flames. As discussed in Mulla \textit{et al.}~\cite{mulla2019interaction}, implementation of these generators for reacting flows is challenging. For non-reacting flow experiments~\cite{hideharu1991realization,mydlarski1998passive,larssen2011generation,bodenschatz2014variable,thormann2014decay,hearst2015effect,neuhaus2020generation,jooss2021spatialb}, the generators are positioned inside wind tunnels with relatively large cross section areas ($\sim1~\mathrm{m}\times\sim1~\mathrm{m}$); however, for reacting flow experiments, the generators need to be integrated with burners with relatively small cross section areas ($\sim0.01-0.1~\mathrm{m}\times\sim0.01-0.1~\mathrm{m}$). This imposes constraints for instrumentation and the largest achievable RMS velocity fluctuations. For example, for the active turbulence generator utilized in Mulla \textit{et al.}~\cite{mulla2019interaction}, the reported $u'$ is 0.26~m/s, which is significantly smaller than $u'$ reported for passive generators. Possibly due to the above limitation, the application of active turbulence generators has not been extensively explored in the area of turbulent combustion. The objective of the present study is to develop and characterize a relatively small size active turbulence generator that allows for producing relatively large RMS velocity fluctuations (compared to those that used active generators for reacting flows) and facilitates stabilization of the relevant turbulent premixed flames. In the present study, the experiments are performed both for the developed active turbulence generator and several passive turbulence generators used in Kheirkhah \textit{et al.}~\cite{kheirkhah2014topology,kheirkhah2015consumption,kheirkhah2016experimental}. Detailed description of the generators and setup, the measurement tools and methods, tested conditions, and the stability of the generated flames are discussed in Section~\ref{section:Exp}. Characteristics of the generated turbulent flow and the conclusions are presented in Sections~\ref{section:results}~and~\ref{section:conc}, respectively.
	
\section{Experimental methodology}
\label{section:Exp}
The details of the burner, the turbulence generation mechanisms, the measurement tools, the tested conditions, as well as the stability of the generated turbulent flames are presented in this section.

\subsection{Burner}
\label{section:burner}
\begin{figure}[!h]
	\centering
	\includegraphics[width=0.95\textwidth]{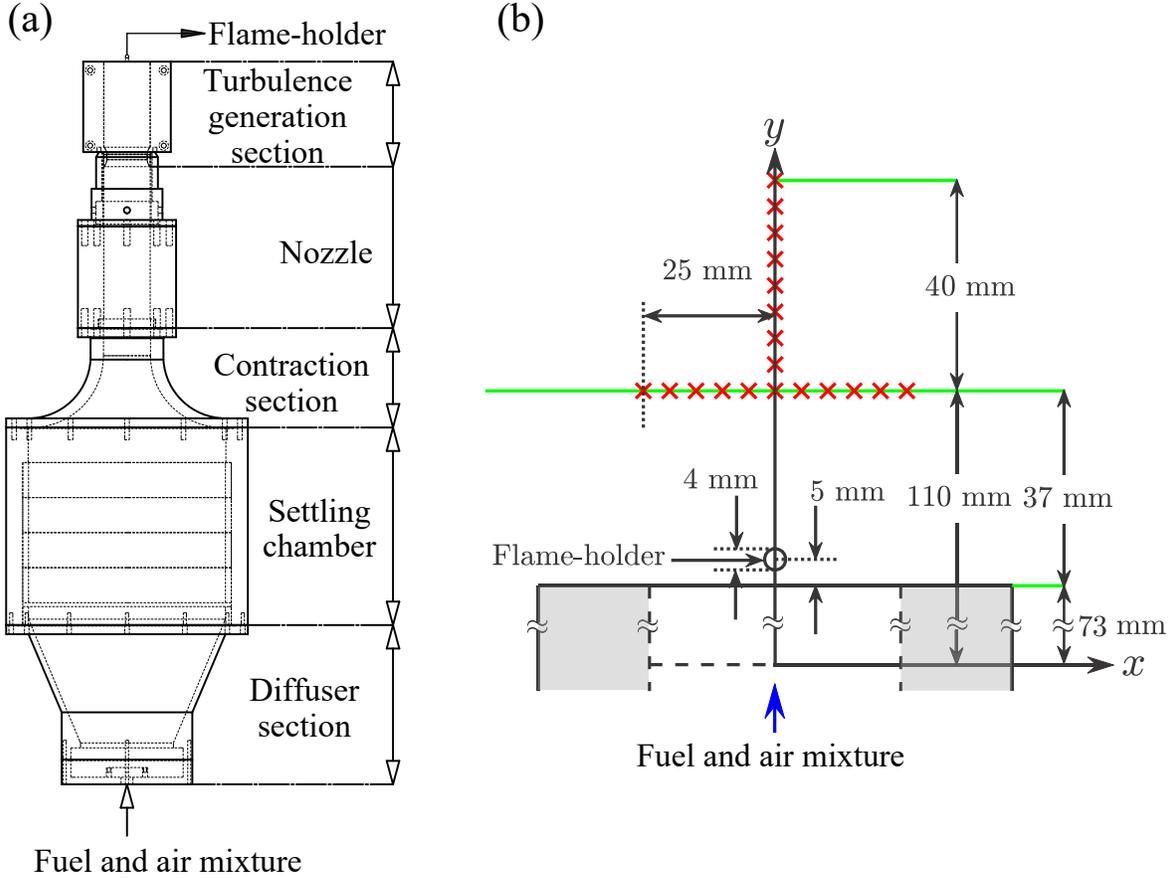}
	\caption{(a) The burner technical drawing and (b) the coordinate system. In (b), the red cross data symbol highlights the locations at which the hot-wire anemometry was performed.}
	\label{fig:setup}
\end{figure}

The burner is similar to that used in the studies of Kheirkhah \textit{et al.}~\cite{kheirkhah2013turbulent,kheirkhah2014influence,kheirkhah2014topology,kheirkhah2015consumption,kheirkhah2016experimental,kheirkhah2016periodic}. The technical drawing of the burner as well as the schematics of the burner exit are shown in Figs.~\ref{fig:setup}(a) and (b), respectively. The burner is composed of a diffuser section with an expansion area ratio of about 4, a settling chamber (which is equipped with equally spaced 5 mesh screens), a contraction section with an area ratio of about 7, a nozzle, and a turbulence generation section. Compared to the studies of Kheirkhah \textit{et al.}~\cite{kheirkhah2013turbulent,kheirkhah2014influence,kheirkhah2014topology,kheirkhah2015consumption,kheirkhah2016experimental,kheirkhah2016periodic} that the turbulence generators are positioned inside the nozzle, in the present study, a turbulence generation section is installed at the exit of the burner and carries the turbulence generating mechanism. As shown in Fig.~\ref{fig:setup}(b), a 4~mm diameter flame-holder is positioned downstream of the turbulence generation section. The distance between the flame-holder centerline and the exit plane of the turbulence generation section was set to 5~mm. A Cartesian coordinate system was used; and, the origin of the coordinate system was positioned at 73~mm upstream of the burner exit plane, see Fig.~\ref{fig:setup}(b). The $y$-- and $z$--axes of the coordinate system are parallel with the nozzle and the flame-holder centerlines, respectively, as shown in Fig.~\ref{fig:setup}(b).

\begin{figure}[!h]
	\centering
	\includegraphics[width=1\textwidth]{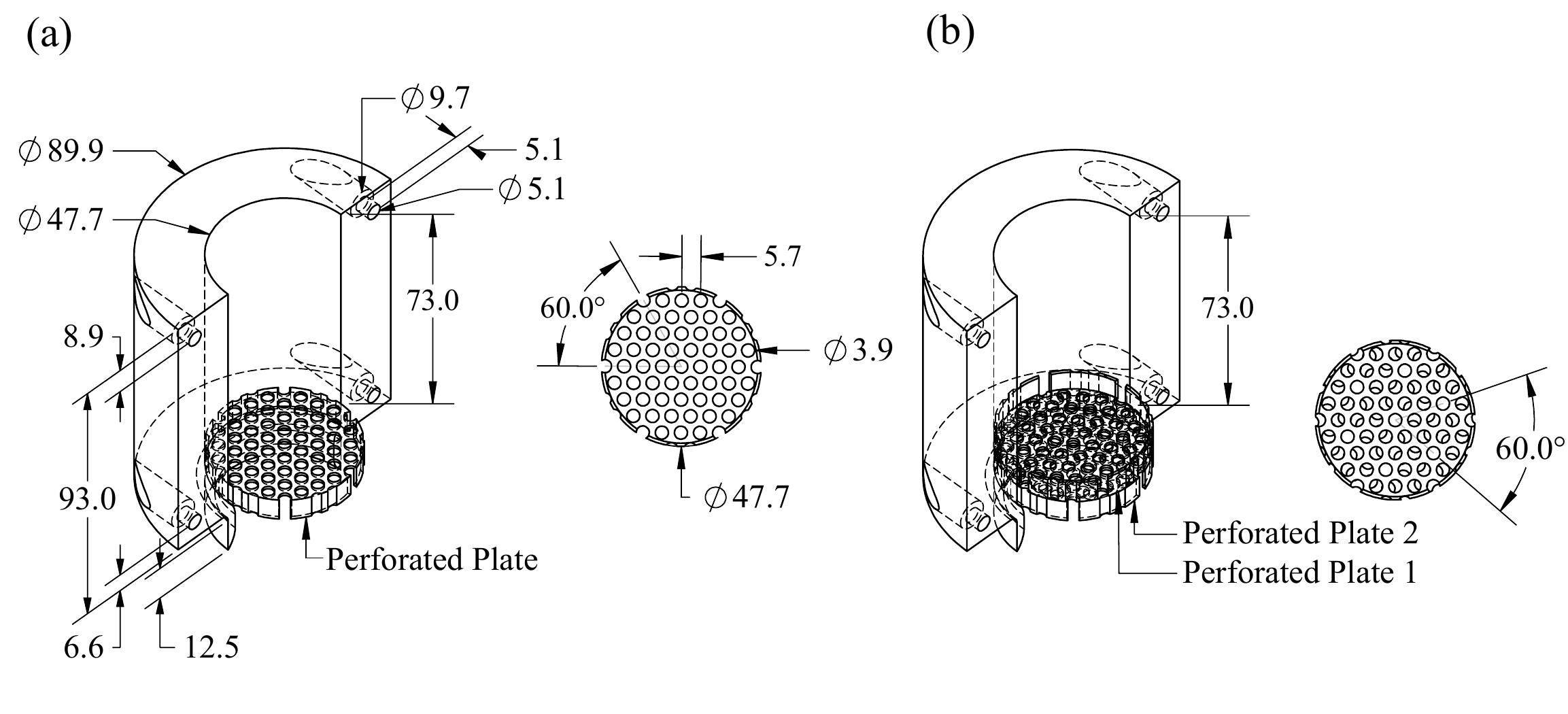}
\caption{The technical drawings of the turbulence generation section equipped with (a) one perforated plate and (b) two perforated plates. The technical drawings of one perforated plate and the orientation of two perforated plates with respect to one another are presented on the right hand sides of (a) and (b), respectively, and are identical to those in Kheirkhah and G\"{u}lder~\cite{kheirkhah2015consumption}.}
	\label{fig:TG}
\end{figure}

Three passive and one active turbulence generation mechanisms were used in the present study. The turbulence generation section shown in Fig.~\ref{fig:setup}(a) was designed to separately carry either of the turbulence generation mechanisms. For passive turbulence generation, either zero, one, or two perforated plates were used. The technical drawings of the passive generators for one and two perforated plates are presented in Figs.~\ref{fig:TG}(a) and (b), respectively. These generators are identical to those used in Kheirkhah and G\"{u}lder~\cite{kheirkhah2015consumption}. The exit plane of one perforated plate and the mid-plane of the combined perforated plates were positioned at 73~mm upstream of the exit plane of the burner, see Figs.~\ref{fig:setup}~and~\ref{fig:TG}. As schematically shown in Fig.~\ref{fig:setup}(b), this upstream location corresponds to 110--150~mm distance between the turbulence generator and the location of measurements shown by the red cross symbols. This distance is about 2.3--3.2 times the turbulence generation section inner diameter. Following the study of Kheirkhah and G\"{u}lder~\cite{kheirkhah2015consumption}, this distance was sufficient for ensuring the measurements locations lie within the turbulence decay region of the perforated plates.

\begin{figure}[!h]
	\centering
	\includegraphics[width=0.7\textwidth]{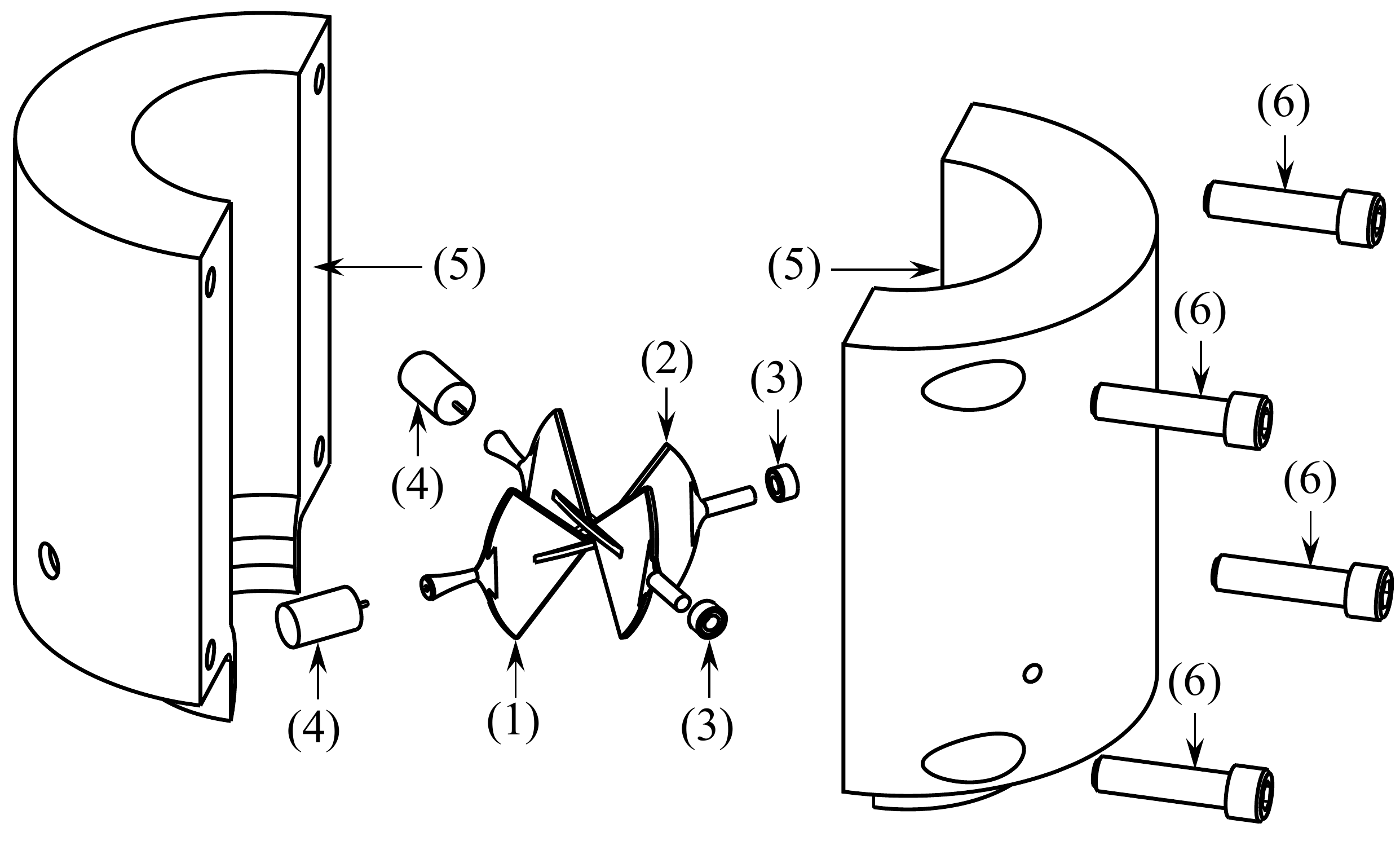}
	\caption{Exploded view of the Compact Flow Stirring Device. Item 1 (Blade 1), item 2 (Blade 2), item 3 (bearings), item 4 (motors), item 5 (casing), and item 6 (fastening screws) are annotated in the figure.}
	\label{fig:assembly}
\end{figure}

\begin{figure}[!h]
	\centering
	\includegraphics[width=1\textwidth]{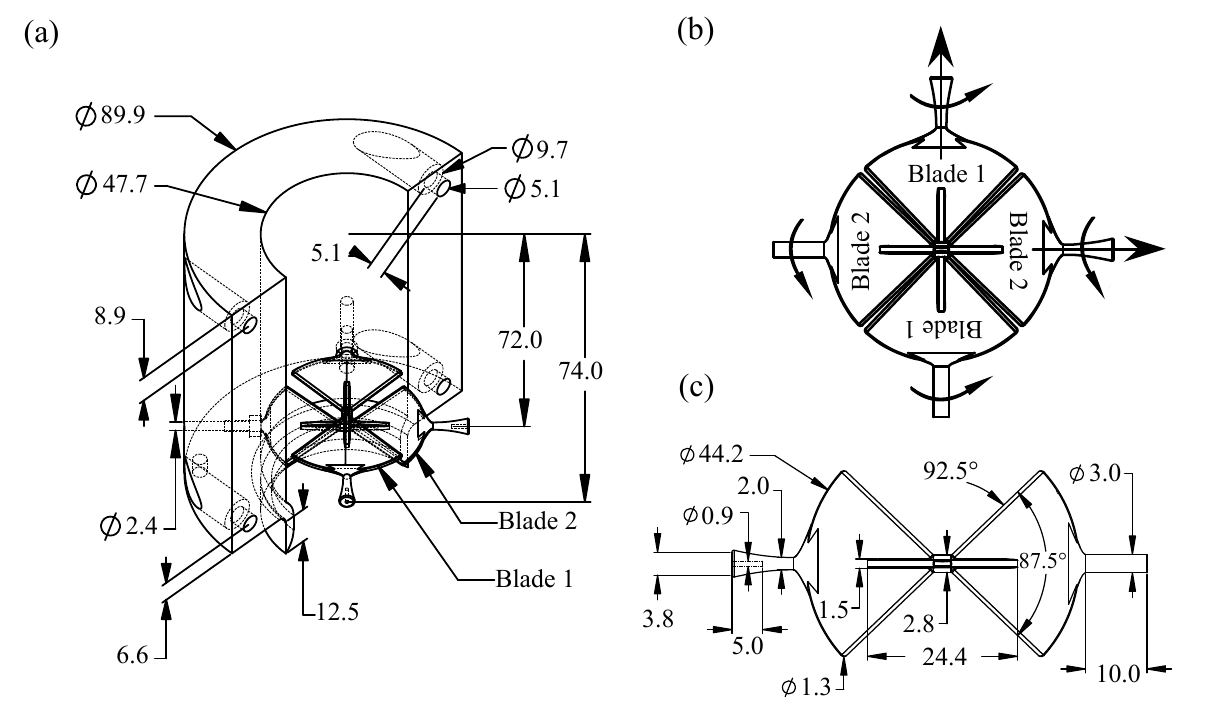}
	\caption{(a) The technical drawings of the turbulence generation section equipped with the Compact Flow Stirring Device. (b) and (c) present both blades positioned normal to the flow direction and the technical drawing of one blade, respectively.}
	\label{fig:CFSD}
\end{figure}

The constituent components of the active turbulence generator, hereafter referred to as the Compact Flow Stirring Device (CFSD), are shown in Fig.~\ref{fig:assembly}. The technical drawing of the CFSD is presented for the first time in the present study. In Fig.~\ref{fig:assembly}, items 1 and 2 are Blades 1 and 2, item 3 is 3~mm inner diameter ball bearings, item 4 refers to the motors, item 5 is the CFSD casing, and item 6 is 1/4 inch-20 Unified Coarse Thread and 1 inch long screws. The technical drawings of items 1, 2, and 5 are shown in Fig.~\ref{fig:CFSD}. Two blades (referred to as Blades 1 and 2) were 3D printed from VeroClear$^{\mathrm{TM}}$, which is made of polymethyl methacrylate. Blades 1 and 2 axes of rotations are both parallel with the exit plane of the turbulence generation section and are positioned 72 and 74~mm upstream of this plane, respectively. Both blades are shown in Fig.~\ref{fig:CFSD}(b), with each blade resembling the shape of a bow-tie. The direction of rotation as well as the axis of rotation for Blades 1 and 2 are shown by the yellow and green colors, respectively. Both blades feature identical technical drawings, with the technical drawing of one blade shown in Fig.~\ref{fig:CFSD}(c). As can be seen, each blade occupies a thin ``pie-shaped" space with an angle of $2\times 87.5^\mathrm{o} = 175^\mathrm{o}$. That is, when positioned normal to the flow direction, both blades occupy $2 \times 175^\mathrm{o}=350^\mathrm{o}$ together. This corresponds to $360^\mathrm{o}-350^\mathrm{o}=10^\mathrm{o}$ opening, which was necessary to avoid blade-blade impact as a result of the blades whirling motion as well as the blades bending due to the moment induced from the salient flow. The area blockage ratio of the blades when positioned parallel to the exit plane of the turbulence generation section is 97\%. Due to the relatively large salient flow static and dynamic loading, the blades are prone to fracture. To minimize the possibility of mechanical failure, two 24.4~mm long and 1.5~mm thick supports were 3D printed on each side of the blades, see Figs.~\ref{fig:CFSD}(b and c).

Compared to the active turbulence generation mechanisms used in past studies relevant to turbulent reacting flows, the Compact Flow Stirring Device (introduced in this study) is relatively light and occupies a small space. These features of the CFSD facilitate extending the application of the active turbulence generators from fundamental combustion science and flows with small velocity fluctuations to relatively more engineering-related problems that feature large velocity fluctuations. Complying with the small weight of the generator, two relatively small size and high-speed motors (2.4G Mini Drone PRO 2) were integrated with the blades for actuation. The motors were controlled using a control system, with details presented in Appendix A.

\subsection{Measurement tools}
\label{section:tools}
The hot-wire anemometry and high-speed imaging were performed to characterize the non-reacting turbulent flow and assess the performance of the CFSD, respectively, with details provided below.

\textit{Hot-wire anemometry}. The hardware for hot-wire anemometry (HWA) includes a probe, a probe support, an sCMOS camera, and two motorized translational stages. The probe is a wire (model 55P01 from Dantec), which is 3~mm long (with an active sensor length of 1.25~mm) and has a diameter of 5~$\mu$m. The wire resistance is 3.58~$\Omega$ at 20$^{\mathrm{o}}$C. A mini-constant temperature anemometry (mini-CTA) circuit (model 9054T0421) maintains the wire temperature, with an overheat ratio of 0.7. A Zyla 5.5 sCMOS camera from Andor which was equipped with a macro lens ($f = 105$~mm) was used to record and ensure that the axis of the probe support (model 9055H0221) was positioned parallel with the main flow direction ($y$--axis). This was performed with an accuracy of $\pm$40~$\mu$m, which is the camera pixel resolution. The motorized translatinal stages (MTS50-Z8 from Thorlabs) featured a 50~mm range of operation. Each motorized translational stage utilized a stepper motor with a resolution of 34555 counts per shaft revolution. Each revolution of the stepper motor shaft led to 1~mm displacement. As a result, the resolution of the stepper motor displacement was 1~mm/34555 = 29~nm. The above range of operation and resolution were sufficient for characterizing the tested turbulent flows. The Kinesis software from Thorlabs was used to position the wire. After the wire was calibrated, with detailed calibration procedure discussed in Appendix B, the measurements were performed. The voltage from the mini-CTA was acquired at $f_\mathrm{s}=10$~kHz. For each tested condition, the HWA data was collected for 90~s, corresponding to 900000 data points.

\begin{figure}[!h]
	\centering
	\includegraphics[width=0.5\textwidth]{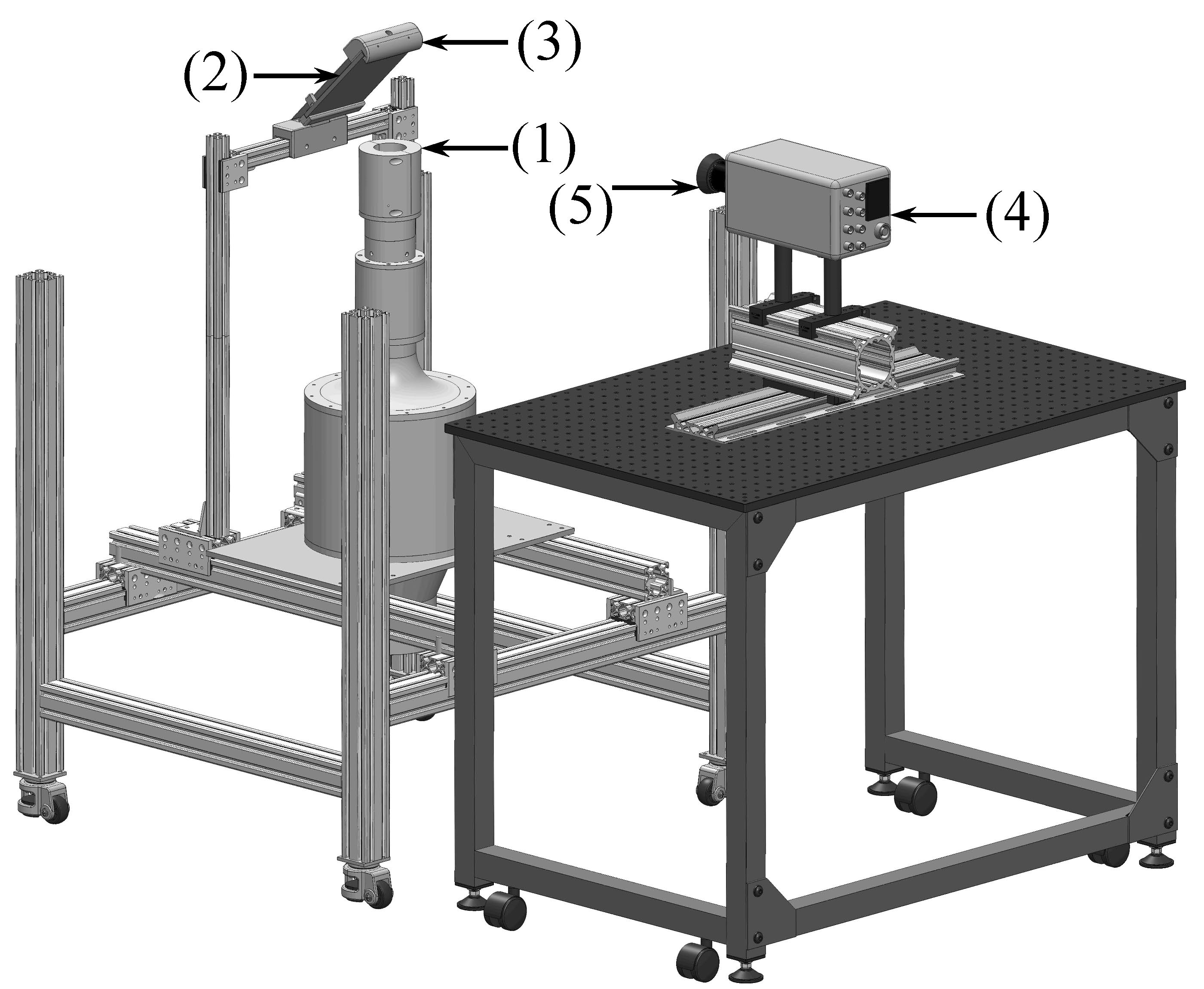}
	\caption{The high-speed imaging setup used for measuring the blades rotation speeds of the Compact Flow Stirring Device. Item 1 (the CFSD), item 2 (flat mirror), item 3 (mirror holder), item 4 (high-speed camera), and item 5 (objective lens) are annotated in the figure.}
	\label{fig:highspeedsetup}
\end{figure}

\textit{High-speed imaging}. This technique was used to measure the rotation speeds of the CFSD blades. Figure~\ref{fig:highspeedsetup} presents the hardware used for this purpose. Specifically, the CFSD (item 1), a flat mirror (item 2), a mirror holder (item 3), a Photron Fastcam S12 camera (item 4), as well as a 50~mm focal length Nikkor lens with aperture size of \#1.8 (item 5) were used, which are shown in the figure. The camera acquisition frequency was set to 2000~Hz, and the images were collected for 5~s.

Using the actuation system presented in Appendix A, equal power was supplied to the motors. Although the blades are nearly identical, possible unequal friction torque can lead to somewhat different blades rotation speeds. It was observed that the robustness of the CFSD operation depends on the salient flow mean bulk velocity. For the tested mean bulk flow velocities, the motors shafts can rotate at a frequency ranging from about 4000 to 6000 revolutions per minute (RPM). It was observed that the most robust operation of the blades are achieved at a rotation speed of 5000~RPM for the tested mean bulk flow velocities. This is the limitation of the CFSD's current design.

\begin{figure}[!h]
	\centering
	\includegraphics[width=1\textwidth]{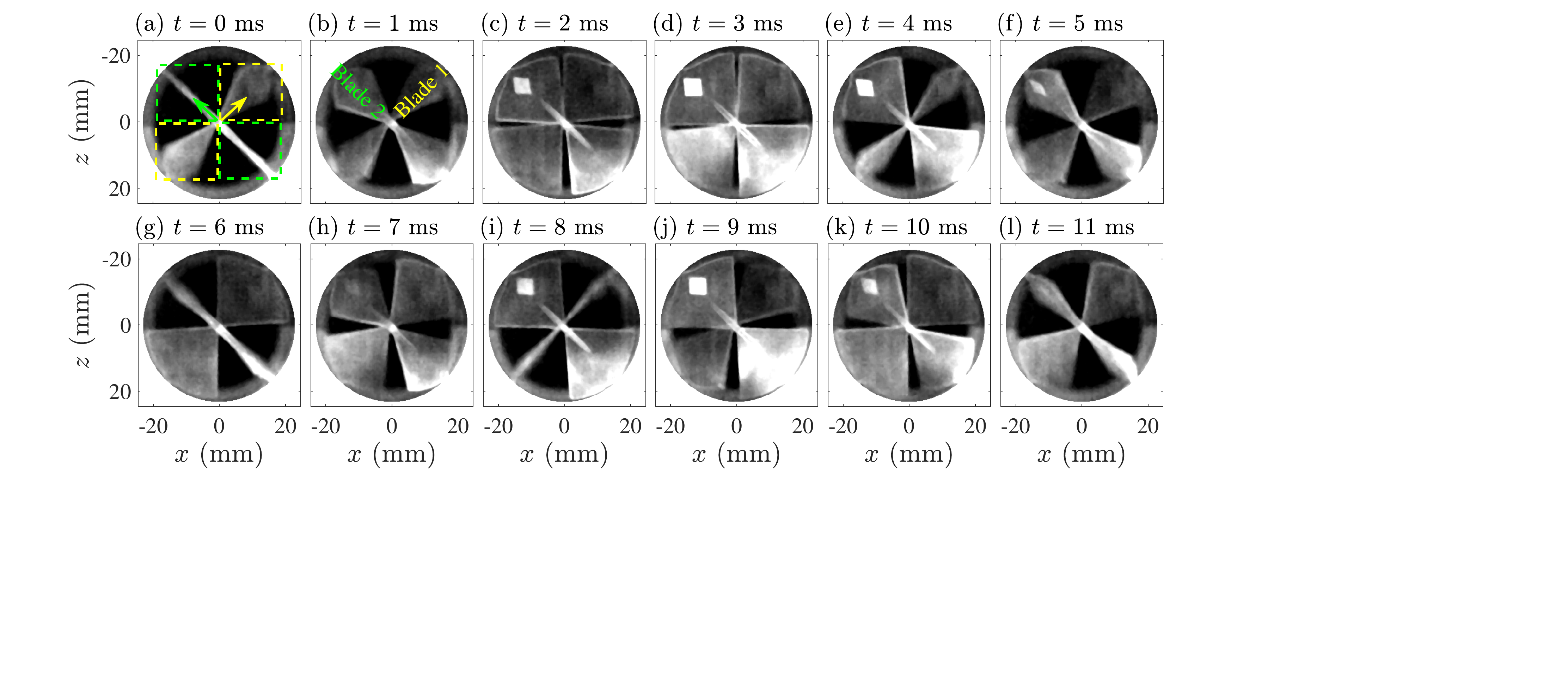}
	\caption{High-speed images of the blades rotation during 11~ms. Every second image is presented for brevity.}
	\label{fig:rotationimages}
\end{figure}

\begin{figure}[!h]
	\centering
	\includegraphics[width=1\textwidth]{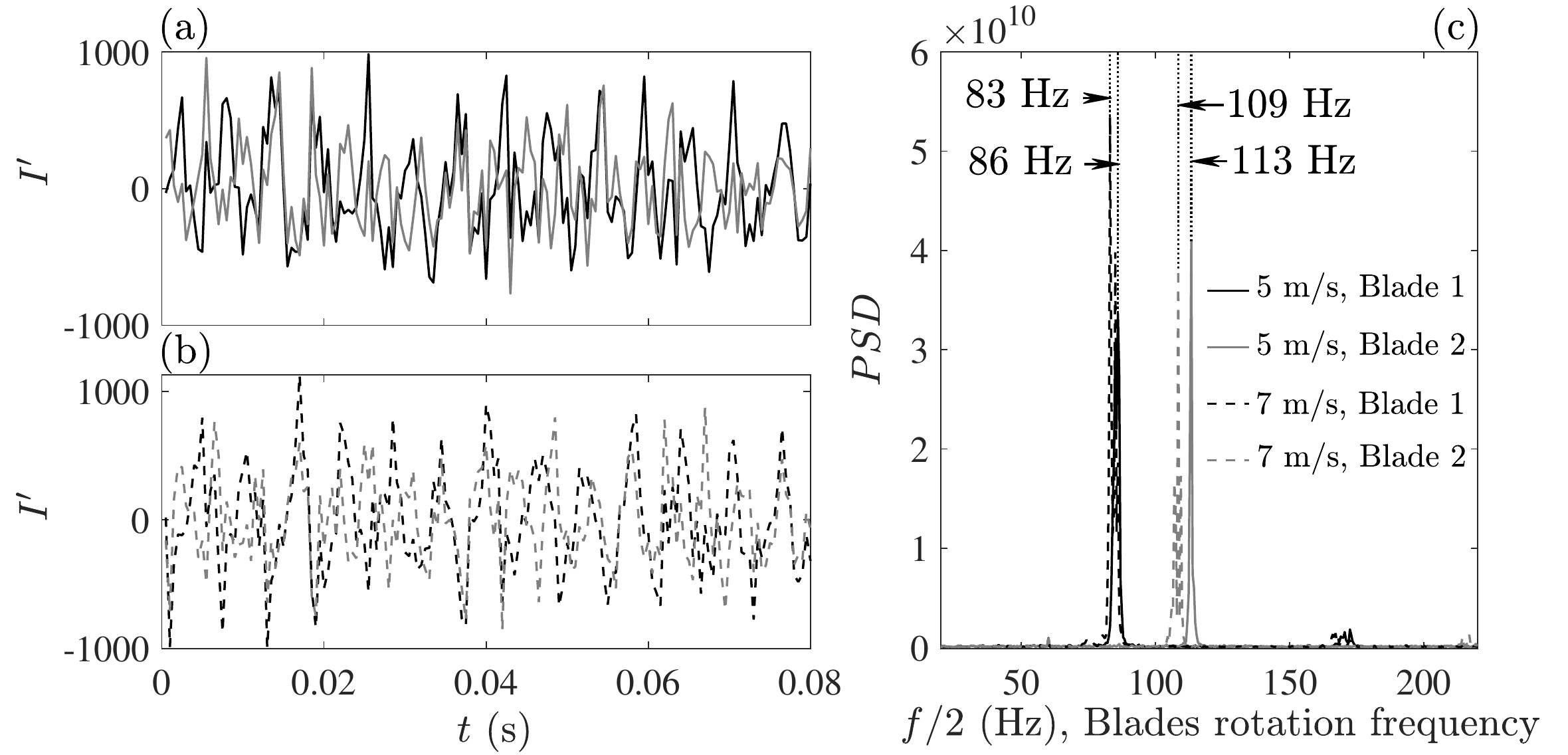}
	\caption{(a and b) are the fluctuations of the spatially averaged light intensity for $U =5.0$ and 7.0~m/s, respectively. (c) is the PSD of $I'$ versus $f/2$.}
	\label{fig:rotationspeed}
\end{figure}

Figure~\ref{fig:rotationimages}(a--l) present the acquired high-speed images for one cycle of Blade 1 rotation. For brevity purposes, every second image is selected for presentation. Blades 1 and 2 axes of rotation are shown by the yellow and green arrows in Fig.~\ref{fig:rotationimages}(a) and correspond to those shown in Fig.~\ref{fig:CFSD}(b). In order to quantify the rotation speed of the blades, the light intensities corresponding to Blades 1 and 2, see the yellow and green boxes in Fig.~\ref{fig:rotationimages}(a), were spatially averaged for each image. The variations of the spatially averaged light intensity subtracted by the corresponding mean (which is referred to as $I'$) versus time are presented in Figs.~\ref{fig:rotationspeed}(a and b) for two mean bulk flow velocities tested in the present study (with details discussed in the next subsection). In Figs.~\ref{fig:rotationspeed}(a and b) the solid and dashed black curves correspond to Blade 1; and, those in solid and dashed gray correspond to Blade 2. As can be seen, variation of $I'$ is nearly periodic for both blades. The Power Spectrum Density (PSD) of $I'$ corresponding to the variations in Figs.~\ref{fig:rotationspeed}(a and b) are presented in Fig.~\ref{fig:rotationspeed}(c). Owing to the blades symmetric geometry with respect to their axes of rotation, $I'$ maximizes twice during one rotation cycle of the blades, suggesting that the frequency of variations in Figs.~\ref{fig:rotationspeed}(a and b) are twice the blades rotation frequency. Thus, the horizontal axis in Fig.~\ref{fig:rotationspeed}(c) is normalized by 2, presenting the rotation frequency of the blades. As can be seen, the rotation frequencies of Blades 1 and 2 are 86~Hz and 113~Hz, respectively, for mean bulk flow velocity of 5.0~m/s. For $U=7.0$~m/s, Blades 1 and 2 rotation frequencies are 83~Hz and 109~Hz, respectively, which are not significantly different from those corresponding to $U=5.0$~m/s.

\subsection{Tested conditions}
\label{section:conditions}
Details of the tested conditions are tabulated in Table~\ref{Tab:Testedconditions}. Four turbulence generation mechanisms (denoted by 0P, 1P, 2P, and CFSD) and two mean bulk flow velocities (5.0 and 7.0~m/s), corresponding to $4\times2 = 8$ non-reacting flow conditions were tested. In the table, $Re_D = UD/\nu$, with $\nu$ being the air kinematic viscosity estimated at 300~K. The air flow rate (hence $U$) was controlled using an Alicat 5000 MCRH. In Table~\ref{Tab:Testedconditions}, $u'$ and $L$ are the RMS of the streamwise velocity fluctuations and the integral length scale estimated based on the streamwise velocity, respectively. Both parameters are calculated for $x = 0$ and $y = 110$~mm, with details discussed in the next section. In the table, the turbulent Reynolds number is calculated from $Re_\mathrm{T} = u'L/\nu$. The Taylor and the Kolmogorov length scales are obtained from $\lambda = LRe_\mathrm{T}^{-1/2}$ and $\eta = LRe_\mathrm{T}^{-3/4}$, respectively. The Reynolds number estimated based on the Taylor length scale ($Re_\lambda$) is calculated from $Re_\lambda = u'\lambda/\nu$. These non-dimensional parameters are tabulated here for completeness and are extensively used in a future study for characterization of the tested turbulent premixed flames. 

\begin{table}[!htbp]
	\caption{Tested conditions of the non-reacting flow. In the table, $L$ is estimated at $x= 0$ and $y = 110$~mm. The unit of $U$ and $u'$ is m/s. The unit of $L$, $\lambda$, and $\eta$ is mm.}
	\label{Tab:Testedconditions}
	\centering
	\scalebox{1.1}{
		\begin{tabular}{ccccccccccc}
			\hline
			\hline
			Symbol & Name & TG & $U$ & $Re_D$ & $u'$ & $L$ & $Re_\mathrm{T}$ & $\lambda$ & $Re_\lambda$ & $\eta$ \\
			\hline
			\textcolor{blue}{\large $\circ$}  &\textcolor{blue}{0P5}&0P&5.0&16000&0.11&9.0&65.8&1.11 & 8.1 &0.39\\
			\hline
			\textcolor{blue}{\large $\bullet$}	&	\textcolor{blue}{0P7}&0P&7.0&22000&0.19&5.9&72.1&0.70& 8.5&0.24\\
			\hline
			\textcolor{green}{$\lozenge$}	&	\textcolor{green}{1P5}&1P&5.0&16000&0.24&4.1&64.2&0.51& 8.0&0.18\\
			\hline
			\textcolor{green}{$\blacklozenge$}&	\textcolor{green}{1P7}&1P&7.0&22000&0.35&4.5&101.8&0.44& 10.1&0.14\\
			\hline
			
			\textcolor{yellow!80!red}{$\square$}	&	\textcolor{yellow!80!red}{2P5}&2P&5.0&16000&0.79&7.2&366.9&0.38& 19.2&0.09\\
			\hline
			\textcolor{yellow!80!red}{$\blacksquare$}	&	\textcolor{yellow!80!red}{2P7}&2P&7.0&22000&1.16&8.4&630.4&0.33& 25.1&0.07\\
			\hline
			\textcolor{red}{$\triangle$}	&	\textcolor{red}{CFSD5}&CFSD&5.0&16000&1.23&10.8&857.1&0.37&29.3 &0.07\\
			\hline
			\textcolor{red}{\large$\blacktriangle$}&	\textcolor{red}{CFSD7}&CFSD&7.0&22000&1.84&9.8&1164.6&0.29& 34.1&0.05\\
			\hline
			\hline
	\end{tabular}}
\end{table}

\subsection{Global stability limits of the turbulent premixed flames generated by the CFSD}
\label{section:stability}
The ultimate goal of the CFSD development and testing is the integration of this device with turbulent premixed combustion equipment. Although the present study is concerned with development of the CFSD and its characterization for non-reacting flows, it is important to demonstrate that for the tested conditions discussed earlier, turbulent premixed flames can be indeed stabilized. While this is discussed here, detailed findings related to the turbulent premixed flames produced by the active turbulence generator is the subject of future work. 

Here, the global stability limits are the minimum and maximum fuel-air equivalence ratios for which the turbulent premixed flames can be stabilized. Tests were conducted for one perforated plate, two perforated plates, and the CFSD and for the mean bulk flow velocities of 5.0 and 7.0~m/s. Experiments were not performed for the test conditions with no perforated plates, as the corresponding flames could stabilize in the settling chamber shown in Fig.~\ref{fig:setup}, which is a safety hazard. Also, global stability of these flames is of least relevance to the industrial applications. Of interest and relevance are two fuels, which are methane and hydrogen-enriched methane. The hydrogen-enriched fuel contains 40\% hydrogen and 60\% methane by volume. The fuel flow rate was controlled using the SLA5853 mass flow controller from Brooks. The fuel and air were mixed inside a premixing chamber, which is positioned about 2~m upstream of the burner. The fuel-air equivalence ratio was varied from 0.35 to 1.00 in 0.05 increments. A Canon VIXIA HF R800 was used to record the broad band luminosity of the tested flames. The camera recorded 60 frames per second, which corresponds to a time separation of about 33~ms between the frames.

\begin{figure}[!h]
	\centering
	\includegraphics[width=1\textwidth]{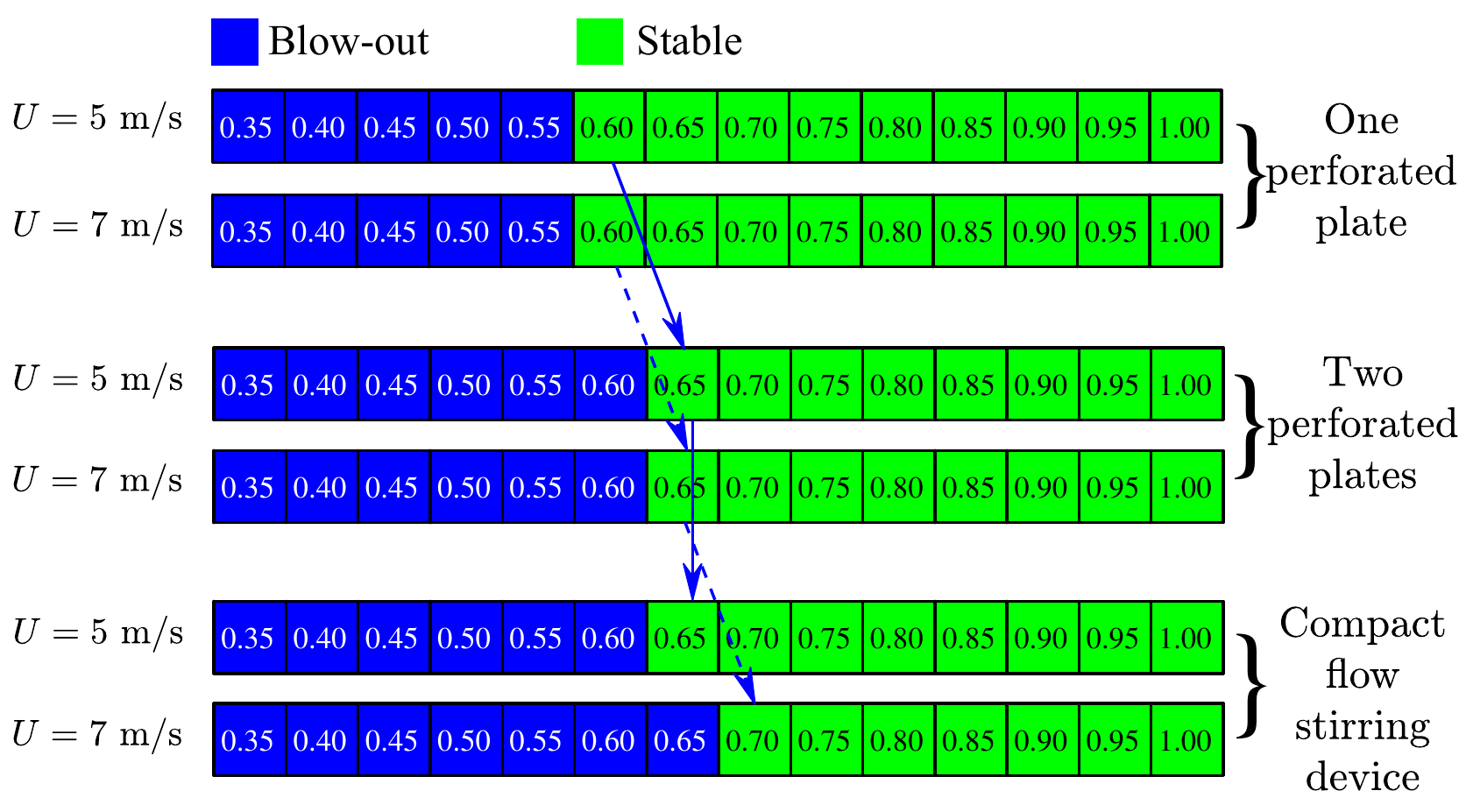}
	\caption{The global stability limits of the tested turbulent premixed methane-air flames. The blue and green colors highlight blow-out and stable conditions, respectively.}
	\label{fig:stability1}
\end{figure}

\begin{figure}[!h]
	\centering
	\includegraphics[width=1\textwidth]{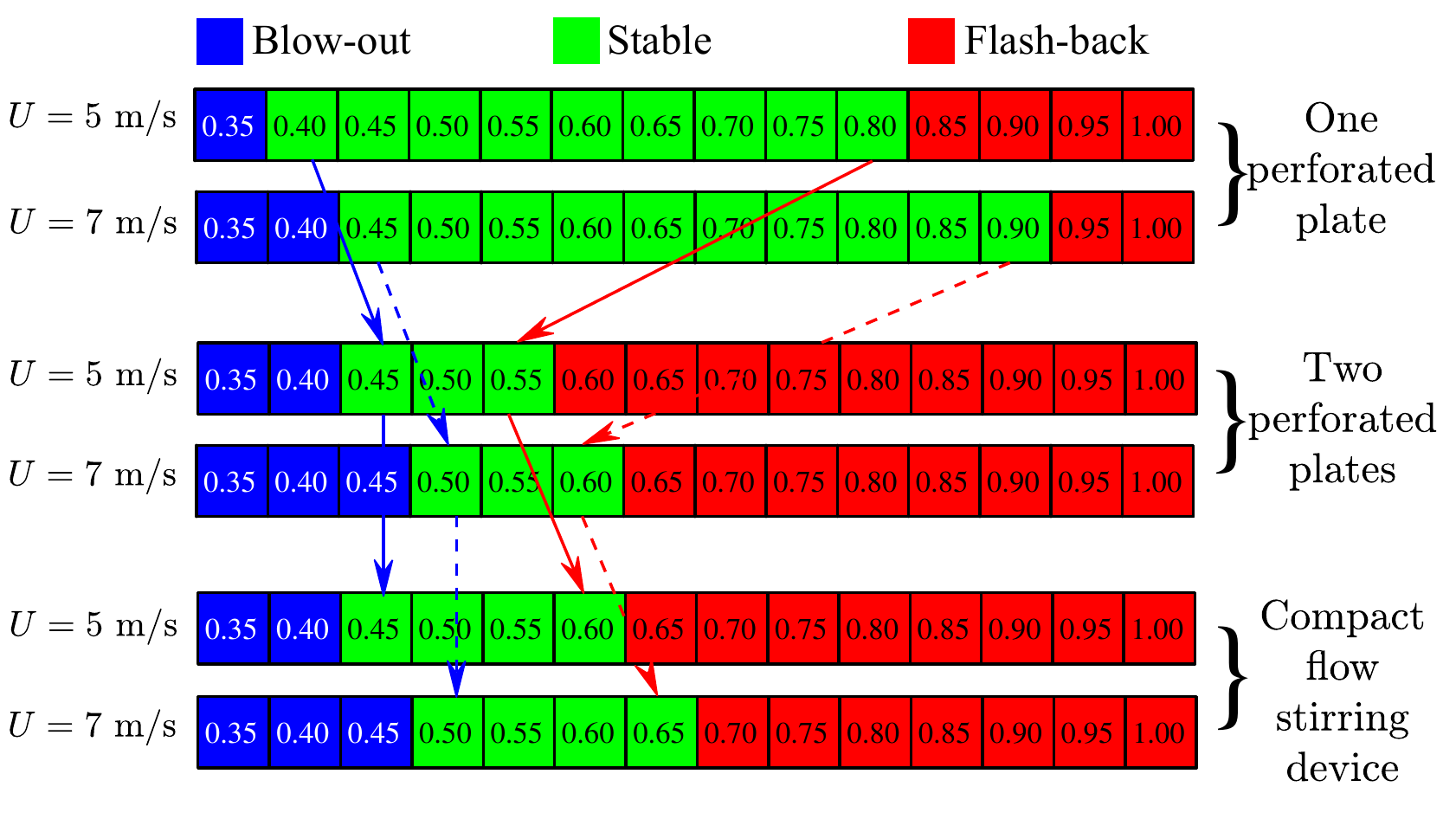}
	\caption{The global stability limits of the tested turbulent premixed 40\% hydrogen-enriched methane-air flames. The blue, green, and red colors highlight the blow-out, stable, and flash-back conditions.}
	\label{fig:stability2}
\end{figure}

The global stability limits for both tested fuels, both mean bulk flow velocities, and the tested turbulence generators were obtained. The results corresponding to methane and hydrogen-enriched methane-air premixed flames are presented in Figs.~\ref{fig:stability1}~and~\ref{fig:stability2}, respectively. In the figures, the blue, green, and red cells pertain to the blow-out, stable, and flash-back conditions, respectively. The results presented in Fig.~\ref{fig:stability1} show that, for turbulent premixed methane-air flames, changing the turbulence generation mechanism from one perforated plate to two perforated plates and to the CFSD increases the fuel-air equivalence ratio near blow-out from 0.60 to 0.65 and 0.60 to 0.70 for $U = 5.0$ and 7.0~m/s, respectively. This is demonstrated using the blue solid and dashed arrows for $U = 5.0$ and 7.0~m/s in Fig.~\ref{fig:stability1}. Similarly, for hydrogen-enriched flames, changing the turbulence generation mechanism from one perforated plate to two perforated plates and to the CFSD increases the fuel-air equivalence ratio near blow-out from 0.40 to 0.45 and 0.45 to 0.50 for $U = 5.0$ and 7.0~m/s, respectively (see the blue arrows in Fig.~\ref{fig:stability2}). Due to the faster combustion chemistry of hydrogen compared to methane, the fuel-air equivalence ratio near blow-out is smaller for the hydrogen-enriched methane-air flames compared to the methane-air flames, which is expected.

Compared to methane-air flames, the hydrogen-enriched methane-air flames featured flash-back for the tested fuel-air equivalence ratios ($0.35\leq \phi \leq 1.00$). For hydrogen-enriched methane-air flames, changing the turbulence generation mechanism from one perforated plate to two perforated plates decreases the near flash-back fuel-air equivalence ratio from 0.80 to 0.55 and 0.90 to 0.60 for $U=5.0$ and 7.0~m/s, respectively, as shown in Fig.~\ref{fig:stability2}. Compared to this observation, however, moving from two perforate plates to the CFSD increases the fuel-air equivalence ratio near flash-back from 0.55 to 0.60 and 0.60 to 0.65 for $U = 5.0$ and 7.0~m/s, respectively (see the red arrows in Fig.~\ref{fig:stability2}).

\begin{figure}[!h]
	\centering
	\includegraphics[width=1\textwidth]{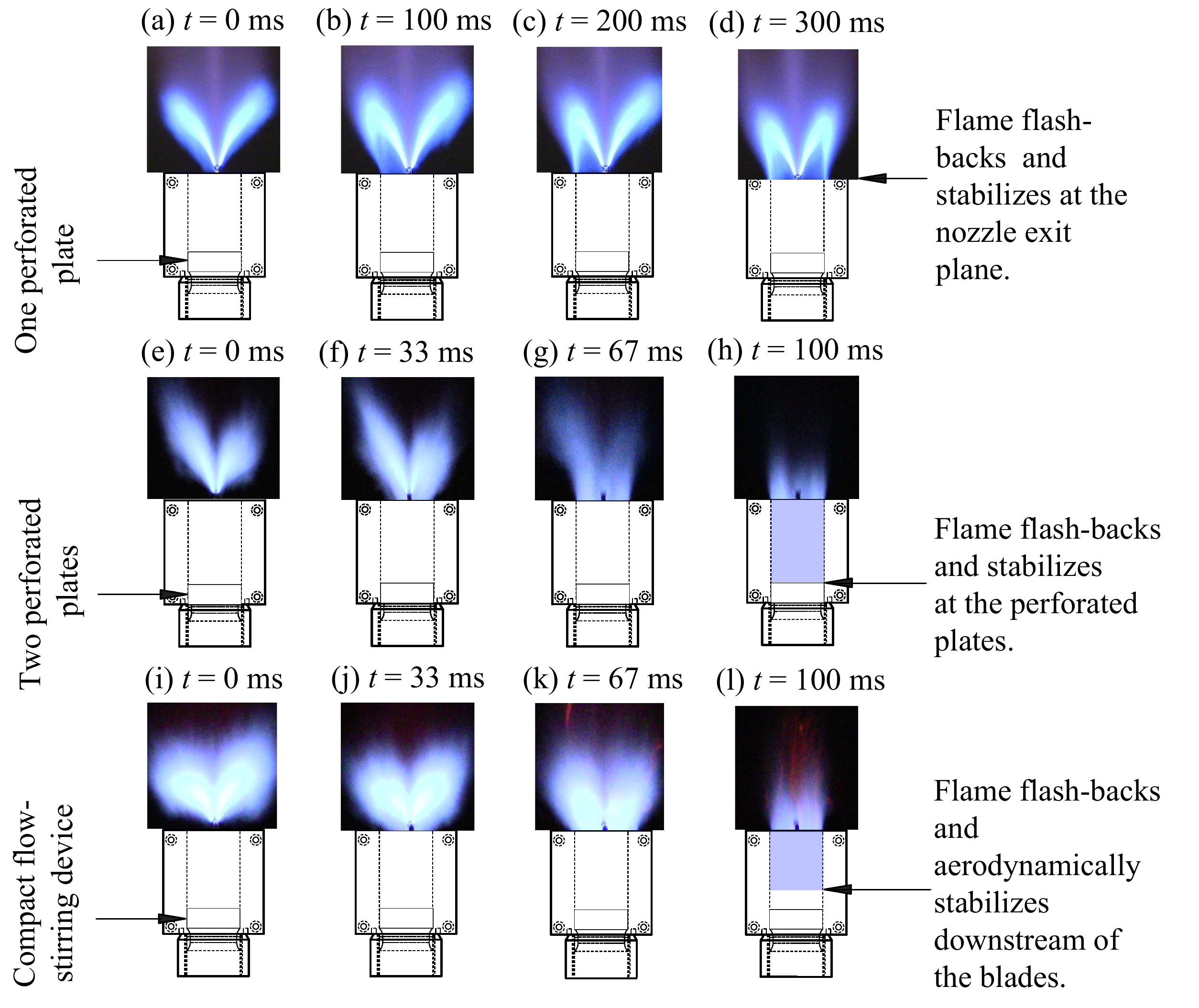}
	\caption{The sequence of images demonstrating the flash-back events for (a--d) one perforated plate, (e--h) two perforated plates, and (i--l) the CFSD. The results correspond to $U=5.0$~m/s.}
	\label{fig:flashback}
\end{figure}

In order to investigate the decreasing and increasing trend of the near flash-back fuel-air equivalence ratio shown in Fig.~\ref{fig:stability2}, the flame natural luminosity was recorded during the flash-back events. The sequence of the images highlighting the flash-back events for one perforated plate, two perforated plates, and the CFSD are shown in Fig.~\ref{fig:flashback}. The images pertain to the mean bulk flow velocity of 5.0~m/s, and similar observations were made for $U=7.0$~m/s. The results show that, first, one wing of the V-shaped flame bends towards upstream and stabilizes at the burner rim during a flash-back event corresponding to one perforated plate, see Figs.~\ref{fig:flashback}(a and b). Then, after about 200~ms, the other flame brush bends and also stabilizes at the burner rim, see Figs.~\ref{fig:flashback}(b--d). This may be due to formation of recirculation zones and velocity deceleration at the burner rim. Compared to the flash-back images for one perforated plate, the results for two perforated plates showed that the flames stabilize farther upstream and at the perforated plates. Compared to two perforated plates, the premixed flames corresponding to the CFSD do not stabilize at the turbulence generator (i.e. the CFSD blades). It was observed that, once flash-back occurred, the flames were aerodynamically stabilized downstream of the CFSD blades and upstream of the flame-holder. This is shown in Figs.~\ref{fig:flashback}(i--l). In essence, changing two perforated plates to the CFSD eliminates stabilization of the flames at the turbulence generator; and, the near flash-back fuel-air equivalence ratio increases for both tested mean bulk flow velocities. These features of the CFSD are of practical importance for industrial applications.

\section{Results}
\label{section:results}

\begin{figure}[!h]
	\centering
	\includegraphics[width=1\textwidth]{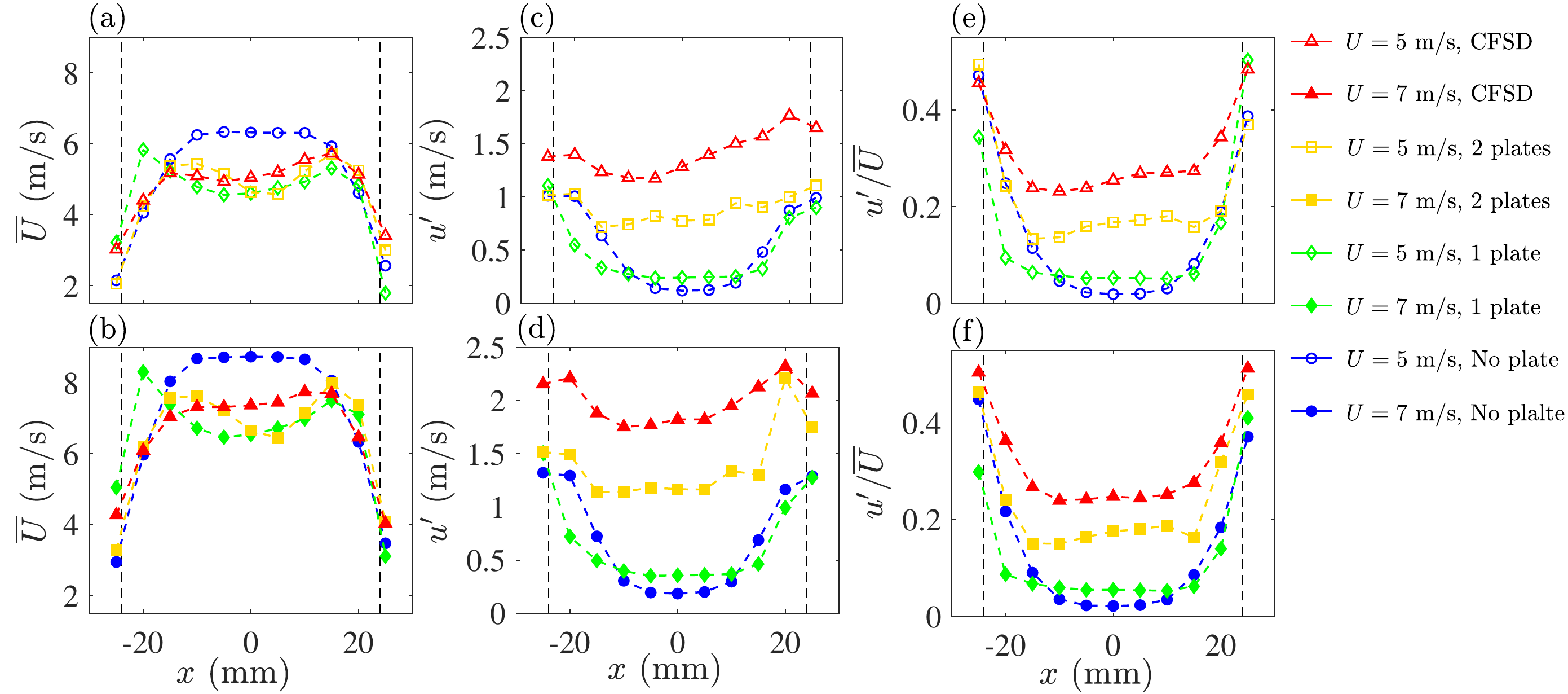}
	\caption{The first, second, and third columns are the mean velocity, RMS velocity fluctuations, and the ratio of these, respectively. The first and second rows correspond to 5.0 and 7.0~m/s tested mean bulk flow velocities.}
	\label{fig:upwo}
\end{figure}

The present study is concerned with two primary features of the flows generated by active grids. First, compared to past wind tunnel relevant studies, it is important for our turbulence generation mechanism to occupy a relatively small space (with widths about that of a gas turbine engine combustor). Second, it is of interest to generate velocity fluctuations that are larger than those of past reacting flow relevant studies that used active turbulence generators. Figure~\ref{fig:upwo} presents the variations of the mean velocity (first column), RMS velocity fluctuations (second column), and the ratio of these two parameters (third column), respectively. The first and second rows correspond to tested mean bulk flow velocities of 5.0 and 7.0~m/s, respectively. As discussed in Appendix B, the uncertainty of the velocity measurements is $\pm$5\% with 95\% confidence level. In Fig.~\ref{fig:upwo}, the data denoted by the blue circular, green diamond, yellow square, and red triangle pertain to no turbulence generator, one perforated plate, two perforated plates, and the Compact Flow Stirring Device, respectively. The vertical dashed lines in the figure highlight the width of the turbulence generation section and is presented for references. The results show that, for the free jet, the mean velocity features a nearly top-hat shaped variation along the $x$--axis; however, those corresponding to the perforated plates as well as the active turbulence generator feature a deceleration close to the burner centerline ($x = 0$~mm). As can be seen in Figs.~\ref{fig:upwo}(c and d), the RMS velocity fluctuations measured at the centerline of the CFSD are 1.2 and 1.8~m/s corresponding to the mean bulk flow velocities of 5.0 and 7.0~m/s, respectively. These RMS velocity fluctuations are about 25\% of the local mean velocity, see Figs.~\ref{fig:upwo}(e and f). As can be seen in these figures, for both mean bulk flow velocities of 5.0 and 7.0~m/s, the CFSD leads to the largest RMS velocity fluctuations and turbulence intensities.

\begin{figure}[!h]
	\centering
	\includegraphics[width=1\textwidth]{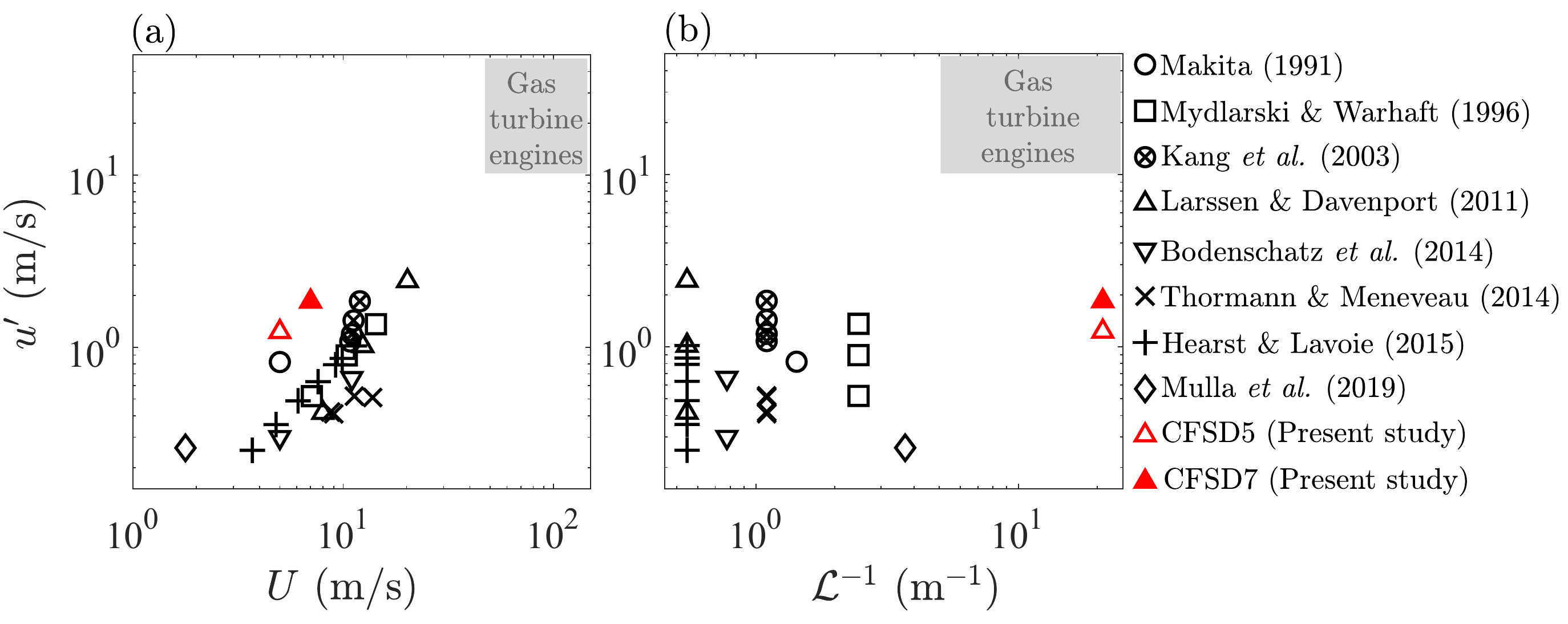}
	\caption{RMS velocity fluctuations versus (a) mean bulk flow velocity and (b) compactness of the active turbulence generator.}
	\label{fig:comparison}
\end{figure}

For comparison purposes, the RMS velocity fluctuations and the mean bulk flow velocities of several investigations~\cite{hideharu1991realization,mydlarski1998passive,larssen2011generation,bodenschatz2014variable,thormann2014decay,hearst2015effect} that used active turbulence generators integrated with wind tunnels, were extracted and presented in Fig.~\ref{fig:comparison}(a). Also, overlaid on the figure are the results of Mulla \textit{et al.}~\cite{mulla2019interaction} which are relevant to turbulent premixed flames and those of the CFSD from the present study. It is acknowledged that $u'$ varies by changing the distance between the turbulence generator and the location of the measurements, and best comparison may be achieved for matching downstream distance normalized by the size of the utilized grid element. Review of past studies suggest this normalized distance varies from a relatively small estimated value of about 6~in Bodenschatz~\textit{et al.}~\cite{bodenschatz2014variable} to about 68 in Mydlarski and Warhaft~\cite{mydlarski1998passive}. Using the radius of the bow-tie shaped flap shown in Fig.~\ref{fig:CFSD}(c) for normalization, our measurements are performed from about 5 to 7 mesh size downstream of the blades, which is relatively small compared to most of the studies listed above. Nonetheless, such location of measurements corresponds to the location of turbulent premixed flame brush and measurements farther downstream were not relevant to the engineering applications of the CFSD.

Comparison of the results from the present study and those from past investigations shows that the CFSD features $u'$ that is significantly larger than that of Mulla~\textit{et al.}~\cite{mulla2019interaction} and many wind tunnel studies that used active generators for non-reacting flows. To assess feasibility (in terms of size) for integration of the generators with gas turbine engine combustors, a parameter referred to as compactness ($1/\mathcal{L}$) is used here. $\mathcal{L}$ is the smallest side length of the tunnel or the exit diameter of the burners. Variation of $u'$ versus $\mathcal{L}^{-1}$ is presented in Fig.~\ref{fig:comparison}(b). For comparison purposes, approximate corresponding parameters for gas turbine engine combustors are also overlaid on Fig.~\ref{fig:comparison} using the gray boxes. As can be seen, $u'$ and $U$ for all active turbulence generators (including the CFSD) are significantly smaller than those relevant to gas turbine engine combustors. However, the compactness of the CFSD is comparable to that of the gas turbine engine combustors. Though larger values of RMS velocity fluctuations can be potentially achieved using the CFSD, excessive vibrations of the blades can impede operation of the device. This design issue will be addressed in the future generations of the CFSD, allowing to achieve larger RMS velocity fluctuations. Nonetheless, the values of $u'$ generated by the CFSD are several folds larger than those of the active turbulence generators used for turbulent premixed combustion studies in the literature.

In the following, a detailed characterization and analysis of the non-reacting flows generated by the CFSD and other tested turbulence generation mechanisms are presented. In Section~\ref{subsection:L}, the turbulent flow length scales are presented. The turbulence decay and the normalized dissipation rate are discussed in Section~\ref{section:decay}. Finally, the spectral content of the turbulent kinetic energy and the dissipation rate are presented in Section~\ref{section:spectra}.

\subsection{Length scales}
\label{subsection:L}
\begin{figure}[!h]
	\centering
	\includegraphics[width=0.6\textwidth]{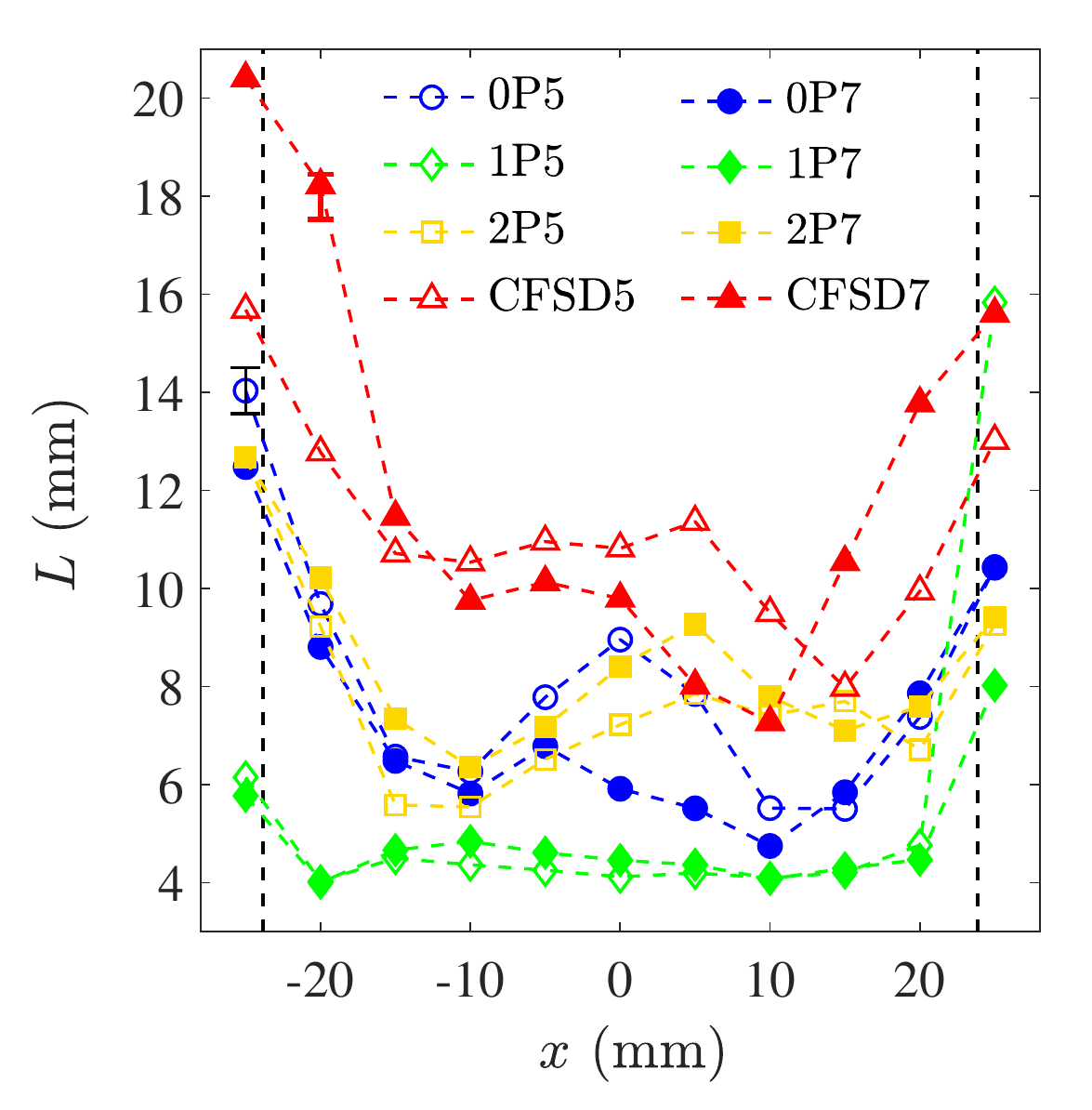}
	\caption{Variation of $L$ along $x$--axis and at $y = 110$~mm. 0P, 1P, 2P, and CFSD correspond to no perforated plate, one perforated plate, two perforated plates, and the Compact Flow Stirring Device, respectively. The open and solid symbols pertain to the mean bulk flow velocities of 5.0 and 7.0~m/s, respectively.}
	\label{fig:ILSx}
\end{figure}

The integral ($L$), Taylor ($\lambda$), and Kolmogorov ($\eta$) length scales were estimated. $L$ was obtained using:

\begin{equation}
	\label{Eq:Ltheory}
	L(x,y,t_0,\Delta T) = U\int_{t_0}^{t_0+t^*} \rho_{uu}(x,y,t_0,\Delta T,\tau) \mathrm{d}\tau,
\end{equation}
where $\rho_{uu}$ is the streamwise velocity auto-correlation and is given by:
\begin{equation}
	\label{Eq:rhotheory}
	\rho_{uu}(x,y,t_0,\Delta T,\tau) = \frac{1}{\Delta T}\int_{t_0}^{t_0+\Delta T}\frac{\tilde{u}(x,y,t)\tilde{u}(x,y,t+\tau)}{\tilde{u}^2(x,y,t)}\mathrm{d}t.
\end{equation}
In Eqs.~(\ref{Eq:Ltheory}~and~\ref{Eq:rhotheory}), $t$ is time; and, $t=0$ is the instant at which the first velocity data is collected. $t_0$ is the starting time at which $\rho_{uu}$ is calculated (thus, $t_0 \geq 0$), $\tau$ is a time delay, $\Delta T$ is the time period of integration for calculation of the streamwise velocity auto-correlation, and $t^*$ is the smallest time at which the auto-correlation of the streamwise velocity equals zero. In Eqs.~(\ref{Eq:Ltheory}~and~\ref{Eq:rhotheory}), $\tilde{u}$ is the streamwise velocity fluctuations with respect to the mean streamwise velocity, i.e. $\tilde{u} = u-\overline{U}$. The integrals in Eqs.~(\ref{Eq:Ltheory}~and~\ref{Eq:rhotheory}) were numerically estimated to calculate $L$. First, a finite number of data points ($N$) was selected to calculate $\Delta T$, which is given by $N = \Delta T f_\mathrm{s}$. Second, a first order central differencing scheme~\cite{lomax2001fundamentals} was used for numerically estimating the integrals in Eqs.~(\ref{Eq:Ltheory}~and~\ref{Eq:rhotheory}). Specifically, $L$ and $\rho_{uu}$ were calculated using

\begin{equation}
	\label{Eq:L}
	L(x,y,i_0,N) \approx \frac{U}{f_\mathrm{s}}\sum_{j = 1}^{j=j^*} \rho_{uu}(x,y,i_0,N,j),
\end{equation}

\begin{equation}
	\label{Eq:rho}
	\rho_{uu}(x,y,i_0,N,j) \approx \frac{1}{N}\sum_{i=i_0+1}^{i=i_0+N}\frac{ \tilde{u}(x,y,t(i))\tilde{u}(x,y,t(i+j-1))}{{\tilde{u}^2(x,y,t(i)))}},	j \in 1:N.
\end{equation}
In Eqs.~(\ref{Eq:L}~and~\ref{Eq:rho}), $i$ and $j$ are summation variables, $i_0 = t_0f_\mathrm{s}$, and $j^* = t^*f_\mathrm{s}$. Analysis of the results suggests that, generally, in addition to the tested condition and the spatial location at which the velocity is measured, the integral length scale can vary by changing $t_0$ (or $i_0$) and $\Delta T$ (or $N$). The influences of $t_0$ and $\Delta T$ on the integral length scale were evaluated and discussed in Appendix C. Variations of the integral length scale at $y = 110$~mm and for $-25 \leq x \leq 25$~mm are presented in Fig.~\ref{fig:ILSx}. The results show that, for all tested conditions and at $y = 110$~mm, moving away from the burner centerline and close to the downstream extension of the interior walls of the burner exit (see the dashed lines in Fig.~\ref{fig:ILSx}) increases the integral length scale. Similar observations are reported in Boxx \textit{et al.}~\cite{boxx2022mechanism} for a jet burner. The smallest reported integral length scale corresponds to one perforated plate at $x = -20$~mm and equals 4~mm. The error bar in Fig.~\ref{fig:ILSx} highlights the largest RMS of $L$ fluctuations obtained by varying $t_0$, as elaborated in Appendix C.

\begin{figure}[!h]
	\centering
	\includegraphics[width=1\textwidth]{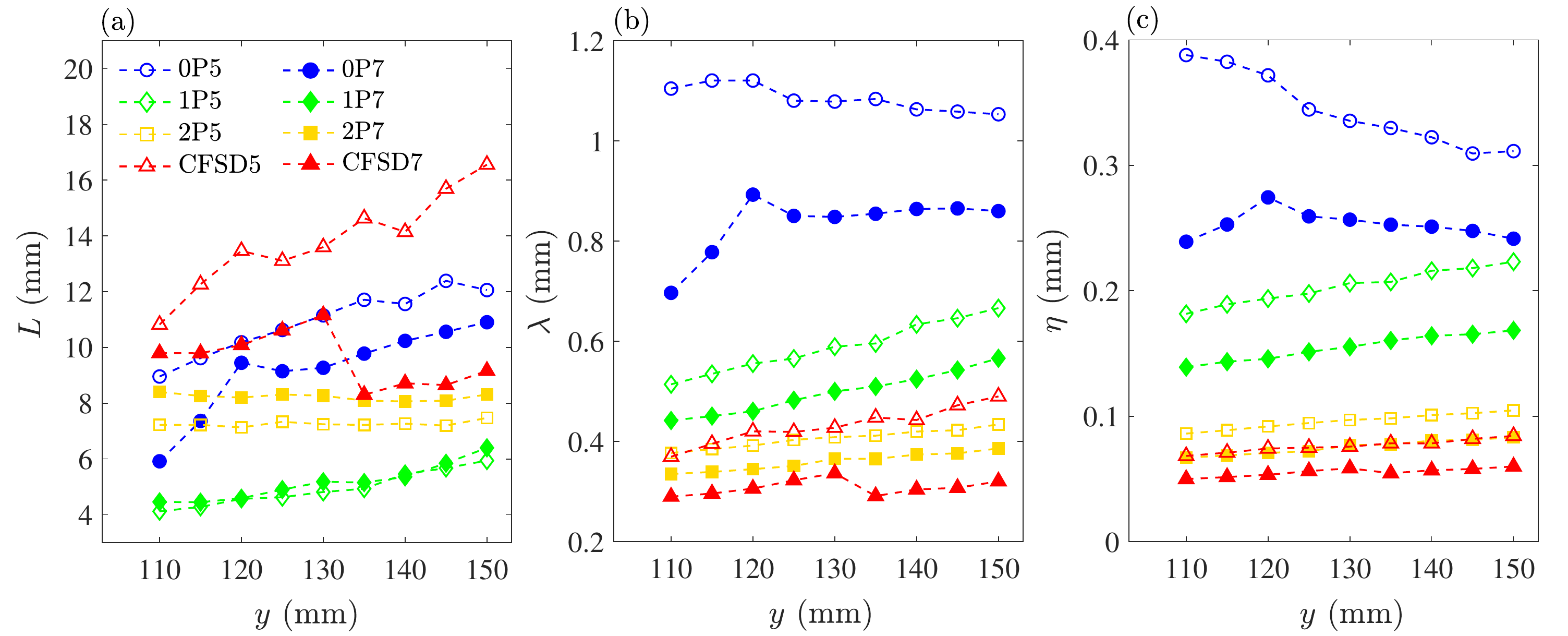}
	\caption{(a) integral, (b) Taylor, and (c) Kolmogorov length scales estimated at $x =0$ and along the vertical axis. 0P, 1P, 2P, and CFSD correspond to no perforated plate, one perforated plate, two perforated plates, and the Compact Flow Stirring Device, respectively. The open and solid symbols pertain to the mean bulk flow velocities of 5.0 and 7.0~m/s, respectively.}
	\label{fig:ILSy}
\end{figure}

\begin{figure}[!h]
	\centering
	\includegraphics[width=1\textwidth]{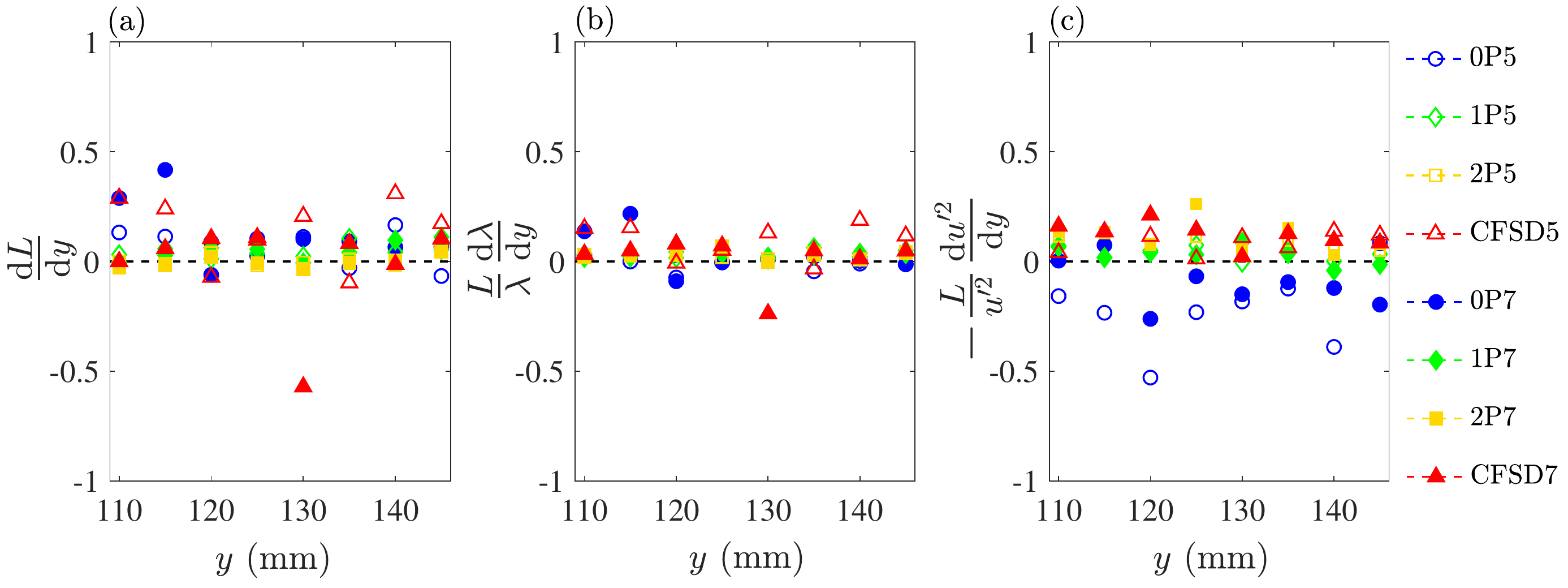}
	\caption{Variations of the LHS of Eqs.~(\ref{Eq:corrsin}a--c) along the vertical axis and at $x = 0$. 0P, 1P, 2P, and CFSD correspond to no perforated plate, one perforated plate, two perforated plates, and the Compact Flow Stirring Device, respectively. The open and solid symbols pertain to the mean bulk flow velocities of 5.0 and 7.0~m/s, respectively.}
	\label{fig:corrsin}
\end{figure}

%\begin{figure}[!h]
%	\centering
%	\includegraphics[width=0.65\textwidth]{Figures/ILS/Auto correlations/rho-modified1.pdf}
%	\caption{The auto-correlation of the streamwise velocity data evaluated at $(x= 0,y = 110~\mathrm{mm})$. For the results presented here, $N = 10^5$ and $i_0 = 1$. The results correspond to the test condition of CFSD7.}
%	\label{fig:rho}
%\end{figure}

Values of $L$, $\Lambda$, and $\eta$ evaluated at $x = 0$ and along the vertical axis are presented in Figs.~\ref{fig:ILSy}(a--c), respectively. The largest measured integral length scale along the burner centerline corresponds to the Compact Flow Stirring Device operating at 5.0~m/s and equals 16.5~mm. The smallest Kolmogorov length scale is 50~$\mu$m and corresponds to the CFSD operating at the mean bulk flow velocity of 7.0~m/s. Corrsin~\cite{corrsin1963turbulence} suggests that, for a homogeneous turbulent flow,

\begin{subequations}
	\label{Eq:corrsin}
	\begin{equation}
		\frac{\mathrm{d}L}{\mathrm{d}y} \ll 1,
	\end{equation}
	
	\begin{equation}
		\frac{L}{\lambda}\frac{\mathrm{d}\lambda}{\mathrm{d}y} \ll 1,
	\end{equation}
	
	\begin{equation}
		-\frac{L}{u'^2}\frac{\mathrm{d}u'^2}{\mathrm{d}y} \ll 1.
	\end{equation}
\end{subequations}
The Left-Hand-Side (LHS) of Eqs.~(\ref{Eq:corrsin}a, b, and c) are presented in Fig.~\ref{fig:corrsin}(a), (b), and (c), respectively. The results in Fig.~\ref{fig:corrsin}(a--c) suggest that the LHS of Eqs.~(\ref{Eq:corrsin}a--c) are generally small and close to zero, suggesting homogeneous turbulent flows are produced by the tested turbulence generators.

\begin{figure}[!h]
	\centering
	\includegraphics[width=1\textwidth]{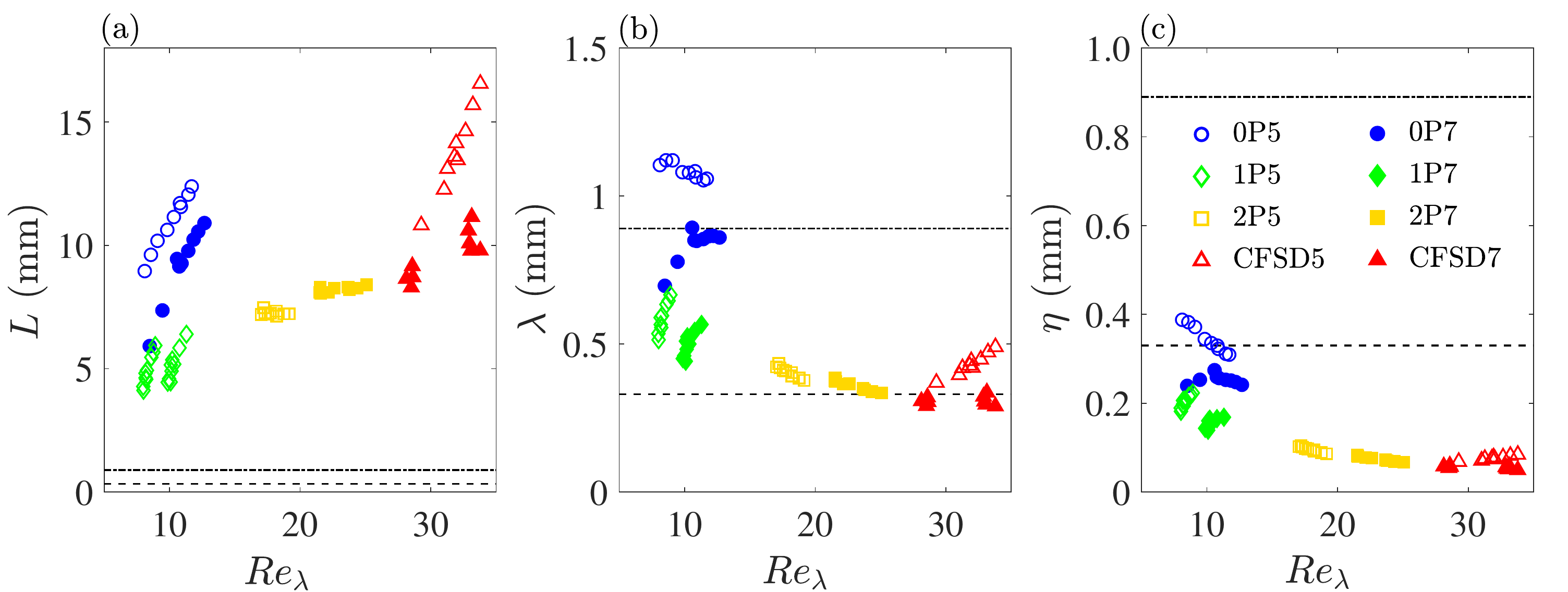}
	\caption{Variations of (a) $L$, (b) $\lambda$, and (c) $\eta$ versus the turbulent Reynolds number estimated based on the Taylor length scale. 0P, 1P, 2P, and CFSD correspond to no perforated plate, one perforated plate, two perforated plates, and the Compact Flow Stirring Device, respectively. The open and solid symbols pertain to the mean bulk flow velocities of 5.0 and 7.0~m/s, respectively. The dashed and dotted-dashed lines are the thicknesses of laminar methane-air (with $\phi = 0.7$) and 40\% hydrogen-enriched methane-air (with $\phi = 0.5$) premixed flames, respectively.}
	\label{fig:ReL}
\end{figure}

Relations between the turbulent flow length scales and a turbulence length scale-based Reynolds number are of interest for studying the structure and burning velocity of the turbulent premixed flames generated by the CFSD, which are the subject of future investigation. G\"{u}lder~\cite{gulder2007contribution} developed mathematical formulations for estimating the contribution of small scale eddies to the burning velocity of turbulent premixed flames using $Re_\lambda$. This formulation was recently modified by the authors~\cite{mohammadnejad2021contributions} as well as Nivarti \textit{et al.}~\cite{nivarti2019reconciling}. For future comparison purposes, $Re_\lambda$ is used here. Variations of $L$, $\lambda$, and $\eta$ versus the turbulent Reynolds number estimated based on the Taylor length scale and are presented in Figs.~\ref{fig:ReL}(a--c). As can be seen, increasing $Re_\lambda$ nearly increases $L$, but decreases $\lambda$ and $\eta$. Although the structure and burning velocity of the turbulent premixed flames generated by the CFSD are the subject of future investigation, the laminar flame thicknesses of the fuels discussed in Section~\ref{section:stability} were estimated~\cite{mohammadnejad2021contributions} and are overlaid on Fig.~\ref{fig:ReL} for comparison purposes. The dashed and dotted-dashed lines pertain to the methane-air (at $\phi = 0.7$) and 40\% hydrogen-enriched methane-air (at $\phi = 0.5$) laminar premixed flame thicknesses, respectively. As can be seen, the Kolmogorov length scale corresponding to the majority of the tested conditions can become smaller than the thickness of both methane-air and hydrogen-enriched methane-air laminar premixed flames. This has implications for the structure and burning velocity of the tested turbulent premixed flames.

\subsection{Turbulence decay and the normalized energy dissipation rate}
\label{section:decay}
\begin{figure}[!h]
	\centering
	\includegraphics[width=0.9\textwidth]{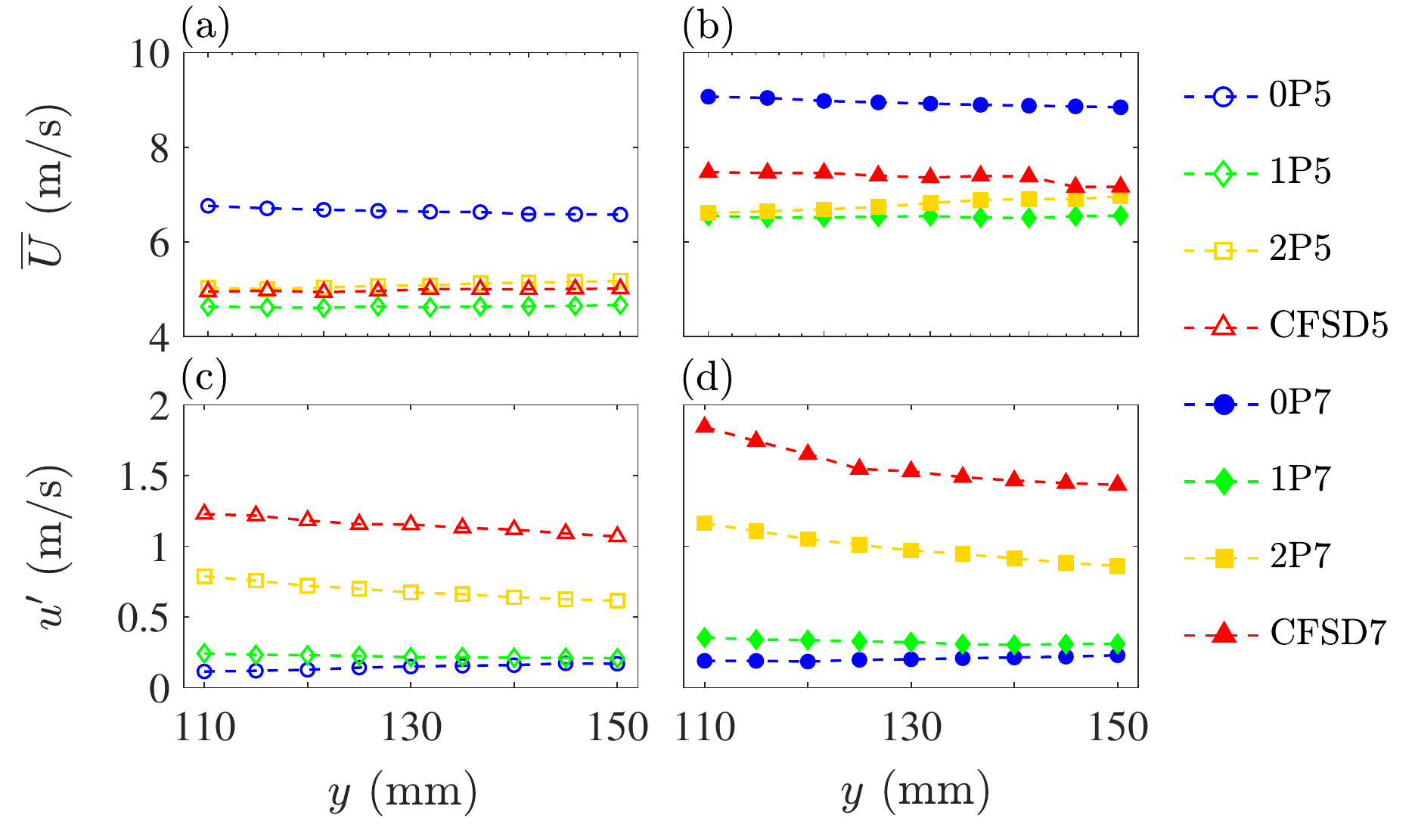}
	\caption{(a and b) are the variations of the mean streamwise velocity along the vertical axis for $U=5.0$~and~7.0~m/s, respectively. (c and d) are the variations of the RMS streamwise velocity fluctuations along the vertical axis for $U=5.0$~and~7.0~m/s, respectively.}
	\label{fig:decay}
\end{figure}

Variation of the mean and RMS velocity fluctuations along the vertical axis and at $x =0$~mm are presented in the first and second rows of Fig.~\ref{fig:decay}. The results presented in the first and second columns pertain to $U=5.0$ and 7.0~m/s, respectively. The data symbols in the figure correspond to those in Table~\ref{Tab:Testedconditions}. For no turbulence generator (the free jet), $\overline{U}$ remains nearly constant for both tested mean bulk flow velocities. This observation is in agreement with the results of Fellouah~\textit{et al.}~\cite{fellouah2009reynolds}. In fact, Fellouah~\textit{et al.}~\cite{fellouah2009reynolds} showed that, for $6000\leq Re_D \leq 30000$ (which corresponds to those tested in the present study, see Table~\ref{Tab:Testedconditions}), $\overline{U}$ remains nearly constant for about 5 diameters downstream of the free jet. Compared to the jet centerline velocity, increasing $y$ from 110~mm to 150~mm increases the turbulence intensity ($u'/\overline{U}$) from about 2\% to 3\% for both tested mean bulk flow velocities.
  
This also agrees with the results presented in Fellouah~\textit{et al.}~\cite{fellouah2009reynolds} for $Re_D$ similar to those tested in the present study. Compared to the test conditions related to no turbulence generator, the results related to the rest of the tested conditions suggest that the RMS velocity fluctuations decreases along the vertical axis.

\begin{figure}[!h]
	\centering
	\includegraphics[width=0.6\textwidth]{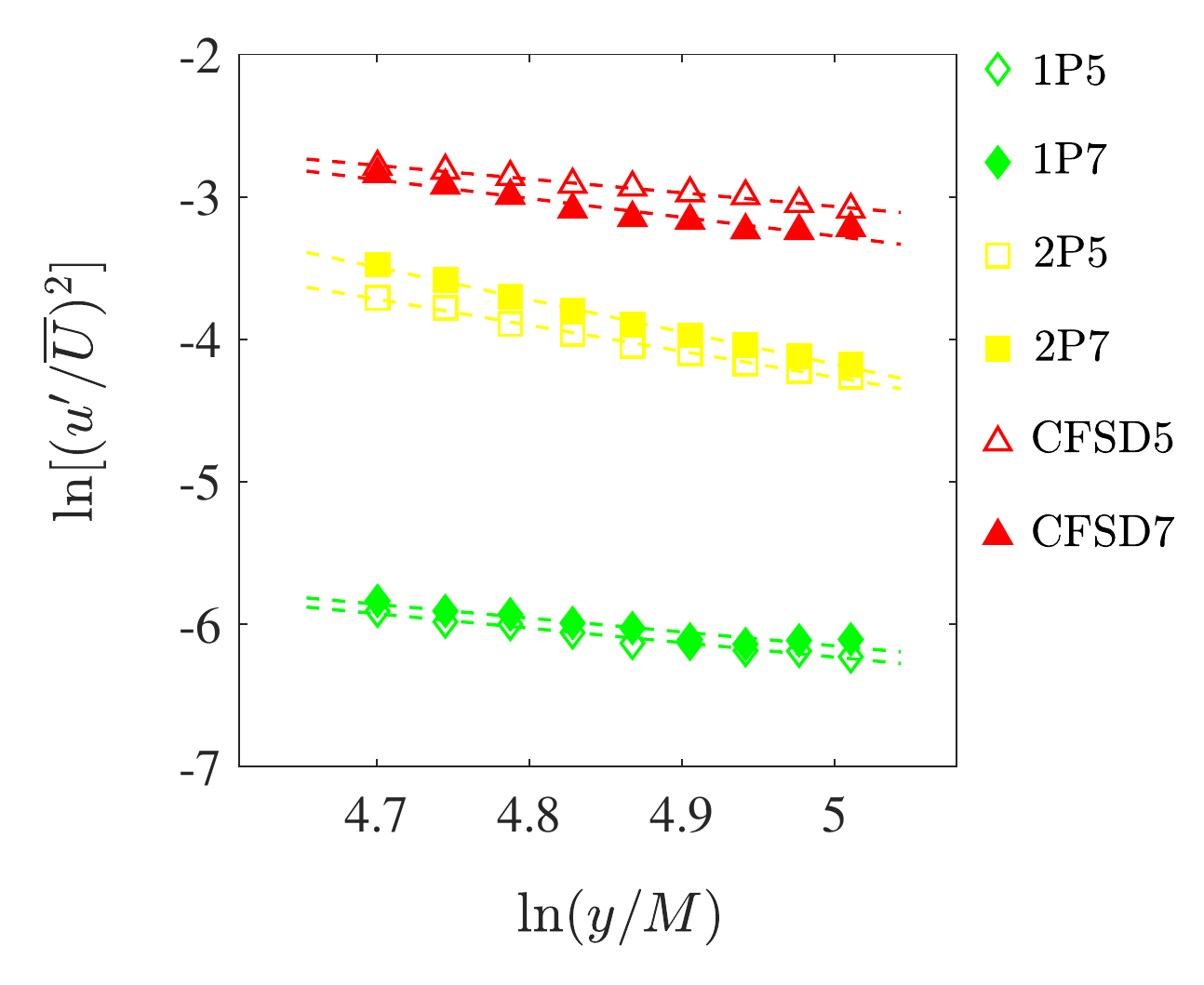}
	\caption{The turbulence decay. The dashed lines present the best linear fit to the data. 1P, 2P, and CFSD correspond to one perforated plate, two perforated plates, and the Compact Flow Stirring Device, respectively. The open and solid symbols pertain to the mean bulk flow velocities of 5.0 and 7.0~m/s, respectively.}
	\label{fig:decaycompiled}
\end{figure}

Since $u'$ decreases with increasing $y$ for one and two perforated plates as well as the Compact Flow Stirring Device, the power-law exponent of the turbulence decay ($n$) was estimated for the corresponding tested conditions. Specifically, $n$ was obtained by calculating the slope of a line fit to the logarithmic variations of the 1D normalized kinetic energy, $(u'/\overline{U})^2$, versus $y/M-y_0/M$ using the least square technique (similar to past investigations~\cite{schedvin1974universal,djenidi2015power,hearst2016effects,hideharu1991realization,mydlarski1998passive,kang2003decaying,thormann2014decay}), which is given by

\begin{equation}
	\label{Eq:powerlaw}
	\ln\left[\left(\frac{u'}{\overline{U}}\right)^2\right] = A+n\ln\left(\frac{y}{M}-\frac{y_0}{M}\right).
\end{equation}
In Eq.~(\ref{Eq:powerlaw}), $A$ is the abscissa of the linear fit, $M$ is the mesh grid size, and $y_0$ is the distance between a ``virtual" origin and the origin of the coordinate system shown in Fig.~\ref{fig:setup}(b). In the present study, $y_0$ was set to 0 and $M$ was selected to be unity. The variations of the 1D normalized turbulent kinetic energy versus $y/M$ as well as the values of $n$ and $A$ are presented in Fig.~\ref{fig:decaycompiled} and Table~\ref{Tab:n}, respectively. Please note that, while the choice of $M$ changes the abscissa of the line fit to the data, $n$ does not depend on $M$. Effect of distance from a virtual origin on $n$ was investigated in the present study as well. It was obtained that such effect is not significant.

\begin{table}[!htbp]
	\caption{Values of $n$ and $A$ in Eq.~(\ref{Eq:powerlaw}). 0P, 1P, 2P, and CFSD correspond to the no perforated plate, one perforated plate, two perforated plates, and the Compact flow Stirring Device, respectively. The numbers in front of the tested conditions labels refer to the tested mean bulk flow velocity.}
	\label{Tab:n}
	\centering
	\scalebox{1.1}{
		\begin{tabular}{ccccccccc}
			\hline
			\hline
			Cond. & \textcolor{blue}{0P5} & \textcolor{blue}{0P7} &\textcolor{green}{1P5} & \textcolor{green}{1P7} & \textcolor{yellow!80!red}{2P5} & \textcolor{yellow!80!red}{2P7} & \textcolor{red}{CFSD5} & \textcolor{red}{CFSD7} \\
			$n$& NA & NA&-1.0&-1.0&-1.8&-2.3&-1.0&-1.3\\
			$A$& NA & NA&-1.2&-1.3&4.9&7.6&1.7&3.9\\
			\hline
			\hline
	\end{tabular}}
\end{table}

\begin{figure}[!h]
	\centering
	\includegraphics[width=0.95\textwidth]{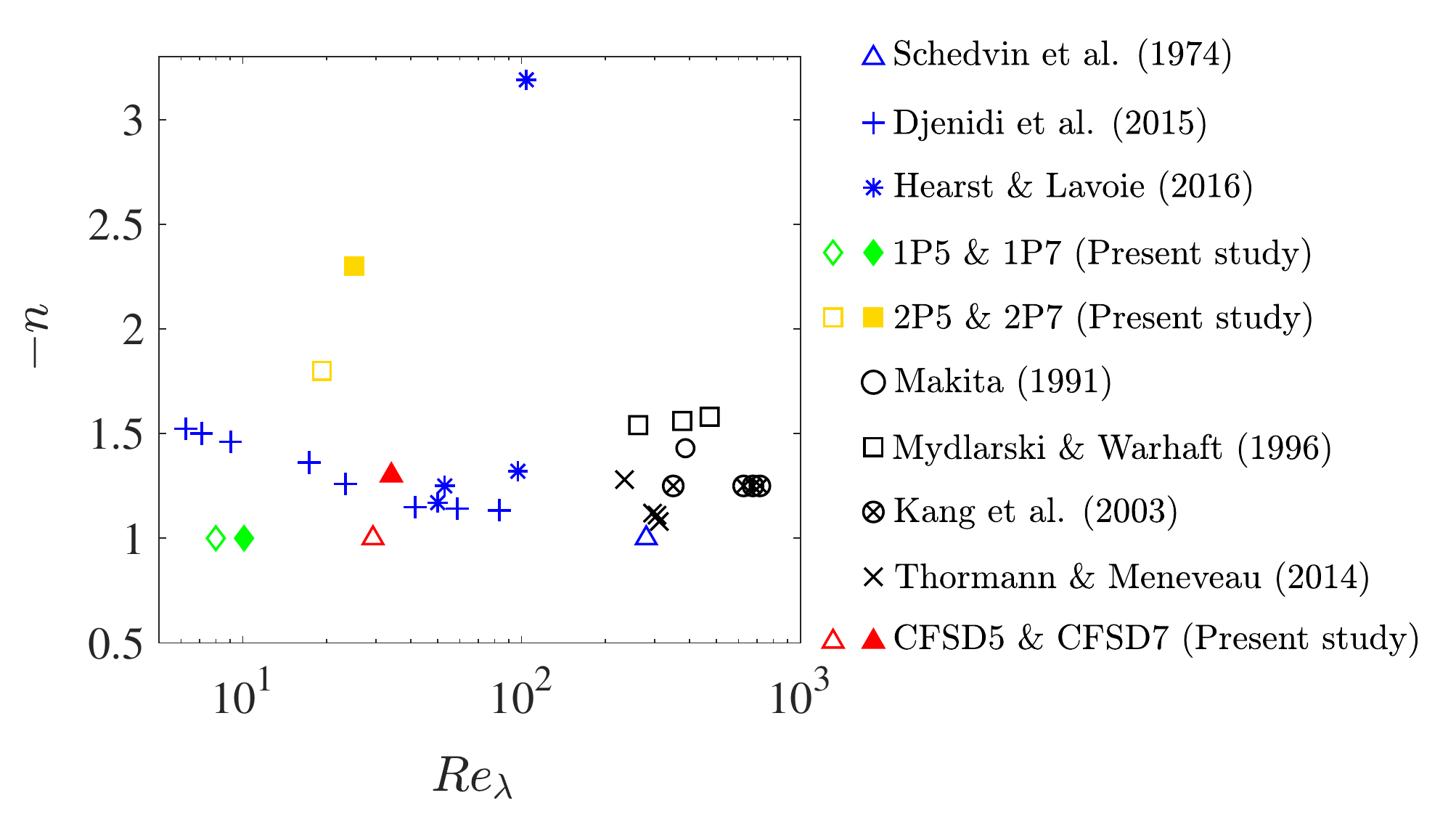}
	\caption{Variation of the negative of the turbulence decay power-law exponent versus the Reynolds number estimated based on the Taylor length scale. The results are compiled from several past investigations~\cite{schedvin1974universal,djenidi2015power,hearst2016effects,hideharu1991realization,mydlarski1998passive,kang2003decaying,thormann2014decay} and the present study. 1P, 2P, and CFSD correspond to one perforated plate, two perforated plates, and the Compact Flow Stirring Device, respectively. The open and solid symbols pertain to the mean bulk flow velocities of 5.0 and 7.0~m/s, respectively.}
	\label{fig:n}
\end{figure}

For homogeneous and isotropic turbulence produced by passive turbulence generators, past studies~\cite{gad1974measurements,lavoie2007effects,hearst2014decay} showed that $n$ is smaller than or equal to -1; and, as suggested by Djenidi \textit{et al.}~\cite{djenidi2015power}, it increases with increasing $Re_\lambda$. Variation of $-n$ versus $Re_\lambda$ is presented in Fig.~\ref{fig:n}. Also overlaid on the figure are the results of past studies~\cite{schedvin1974universal,djenidi2015power,hearst2016effects,hideharu1991realization,mydlarski1998passive,kang2003decaying,thormann2014decay}. The blue, green, and yellow data points correspond to the results from the literature that used passive turbulence generators, 1P5\&7, and 2P5\&7, respectively. The black and red data points correspond to the studies that used active turbulence generators and CFSD5\&7, respectively. Compared to that suggested in Djenidi \textit{et al.}~\cite{djenidi2015power}, our analysis does not show that $n$ follows a relation with $Re_\lambda$ for either passive or active turbulence generators. The largest reported value of $-n$ corresponds to that of Hearst and Lavoie~\cite{hearst2016effects}, which is 3.2, and Mydlarski and Warhaft~\cite{mydlarski1998passive}, which is 1.6, for passive and active turbulence generators, respectively.

\begin{figure}[!h]
	\centering
	\includegraphics[width=0.8\textwidth]{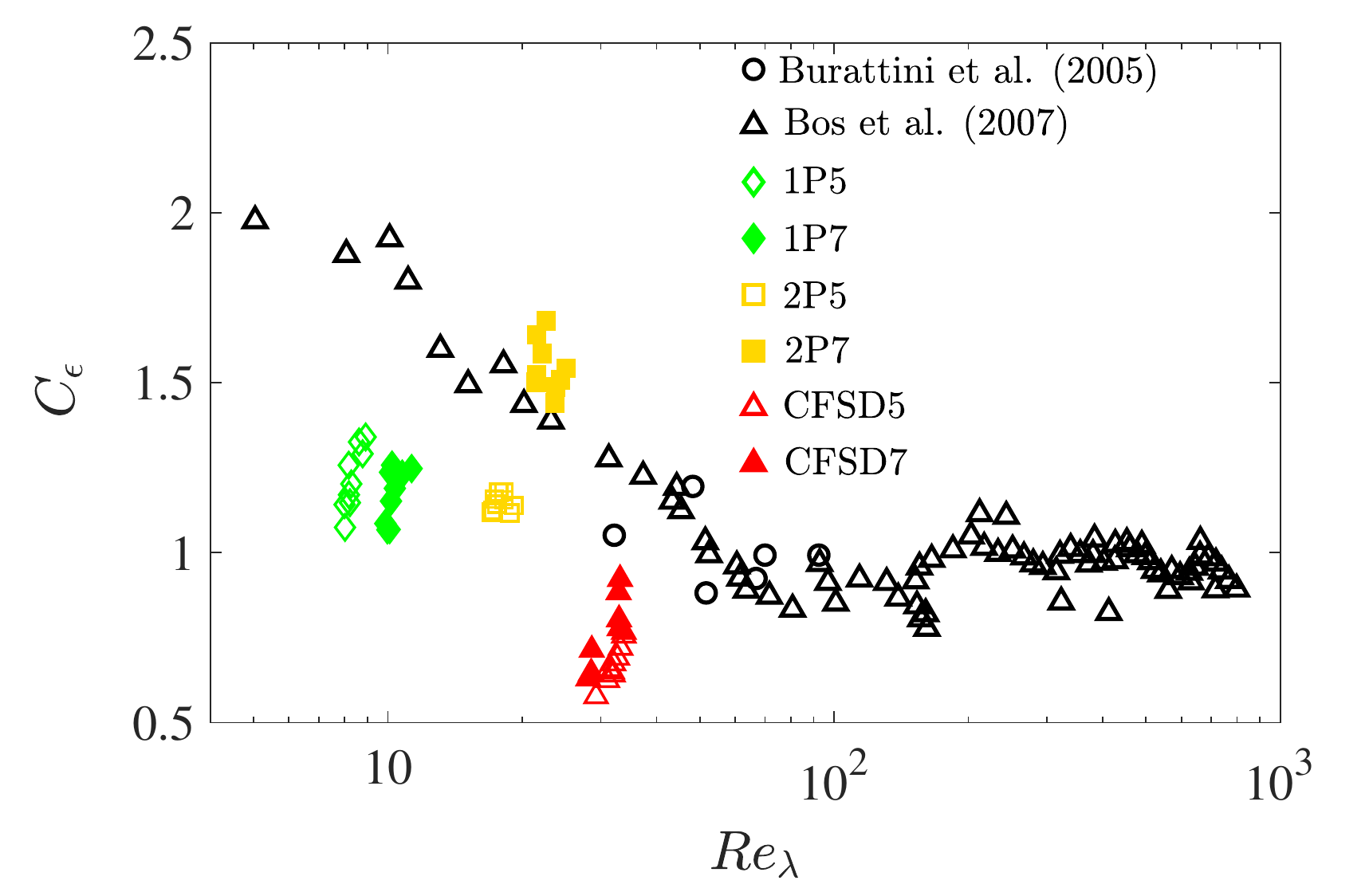}
	\caption{The normalized energy dissipation rate for the tested conditions that the turbulence decays along the vertical axis. The results of Burattini~\textit{et al.}~\cite{burattini2005normalized}~and~Bos~\textit{et al.}~\cite{bos2007spectral} are shown by the black circular and triangular data symbols, respectively, for comparison purposes. 1P, 2P, and CFSD correspond to one perforated plate, two perforated plates, and the Compact Flow Stirring Device, respectively. The open and solid symbols pertain to the mean bulk flow velocities of 5.0 and 7.0~m/s, respectively.}
	\label{fig:dissipation}
\end{figure}

\begin{figure}[!h]
	\centering
	\includegraphics[width=0.7\textwidth]{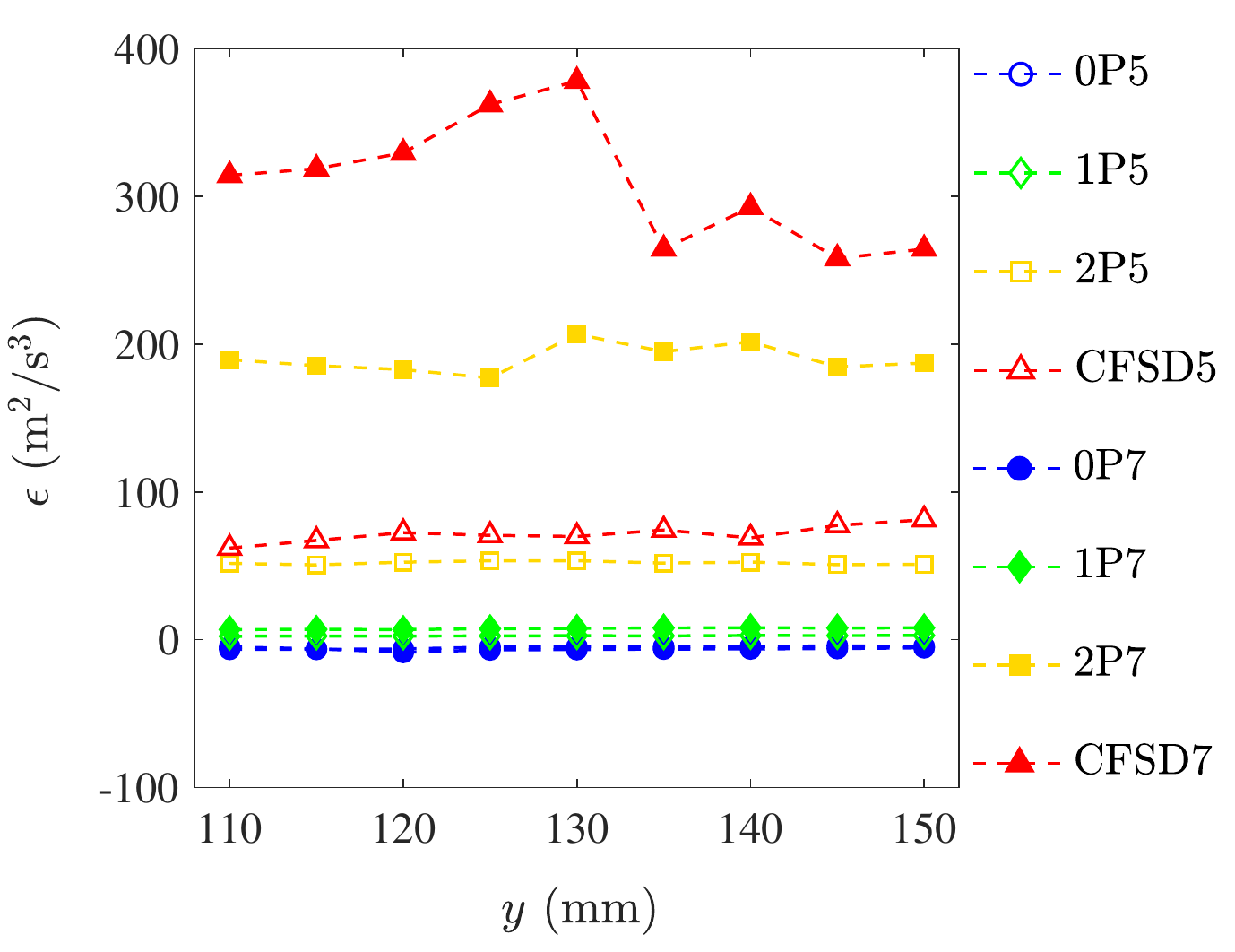}
	\caption{Variation of the dissipation rate versus $y$. 0P, 1P, 2P, and CFSD correspond to no perforated plate, one perforated plate, two perforated plates, and the Compact Flow Stirring Device, respectively. The open and solid symbols pertain to the mean bulk flow velocities of 5.0 and 7.0~m/s, respectively.}
	\label{fig:dissipationrate}
\end{figure}

Variations of the turbulent kinetic energy and the integral length scale along the vertical axis can be used to estimate the normalized energy dissipation rate ($C_\epsilon$). For large values of $Re_\lambda$, the normalized energy dissipation rate is calculated from Burattini~\textit{et al.}~\cite{burattini2005normalized}

\begin{equation}
	C_\epsilon = \frac{L\epsilon}{u'^3},
\end{equation}
where the dissipation rate is given by
\begin{equation}
	\epsilon = \frac{-\overline{U}}{2}\frac{\mathrm{d}q'^2}{\mathrm{d}y},
	\label{Eq:epsilon}
\end{equation}
and $q'^2$ is the summation of all velocity components variances. Estimation of $q'^2$ requires measurements of all three component of the velocity, which was not performed in the present study. For an isotropic flow, $q'^2 = 3u'^2$. This relation was used to estimate $q'^2$, and it is acknowledged that deviation from isotropy can influence the below analysis. Additionally, it was assumed that the power-law decay of the 1D normalized turbulent kinetic energy is given by Eq.~(\ref{Eq:powerlaw}), which is an acceptable assumption following the collapse of the results shown in Fig.~\ref{fig:decaycompiled}. Equations~(\ref{Eq:powerlaw}--\ref{Eq:epsilon}) can be combined to show that

\begin{equation}
	\label{Eq:Ce}
	C_\epsilon \approx \frac{-3e^An\overline{U}^3}{2u'^3}\left(\frac{y}{M}-\frac{y_0}{M}\right)^{n-1}.
\end{equation}
Variation of $C_\epsilon$ versus $Re_\lambda$ corresponding to the test conditions that the turbulence decays along the vertical axis as well as those of Burattini \textit{et al.}~\cite{burattini2005normalized}~and~Bos~\textit{et al.}~\cite{bos2007spectral} are presented in Fig.~\ref{fig:dissipation}. Both of these studies show that increasing $Re_\lambda$ nearly decreases $C_\epsilon$. The result of the present study shown in Fig.~\ref{fig:dissipation} suggest that, for a fixed mean bulk flow velocity, the values of the normalized energy dissipation rate estimated for the CFSD are smallest among the tested turbulence generators. It is important to note that the small values of $C_\epsilon$ reported for the CFSD do not suggest that $\epsilon$ is also necessarily small for this turbulence generator. In order to investigate this, variation of the energy dissipation rate versus the vertical distance is presented in Fig.~\ref{fig:dissipationrate}. Indeed, as can be seen, for a fixed tested mean bulk flow velocity, the CFSD features the largest $\epsilon$ among the tested turbulence generators. In the next subsection, the spectral analysis is used to investigate this in detail.

%The decreasing trend for the relation between $C_\epsilon$ and $Re_\lambda$ of past studies \cite{burattini2005normalized,bos2007spectral} is anticipated for relatively large values of $Re_\lambda$, which is not the case for the tested conditions of the present study.

\subsection{Turbulent kinetic energy and dissipation spectra}
\label{section:spectra}
\begin{figure}[!h]
	\centering
	\includegraphics[width=0.9\textwidth]{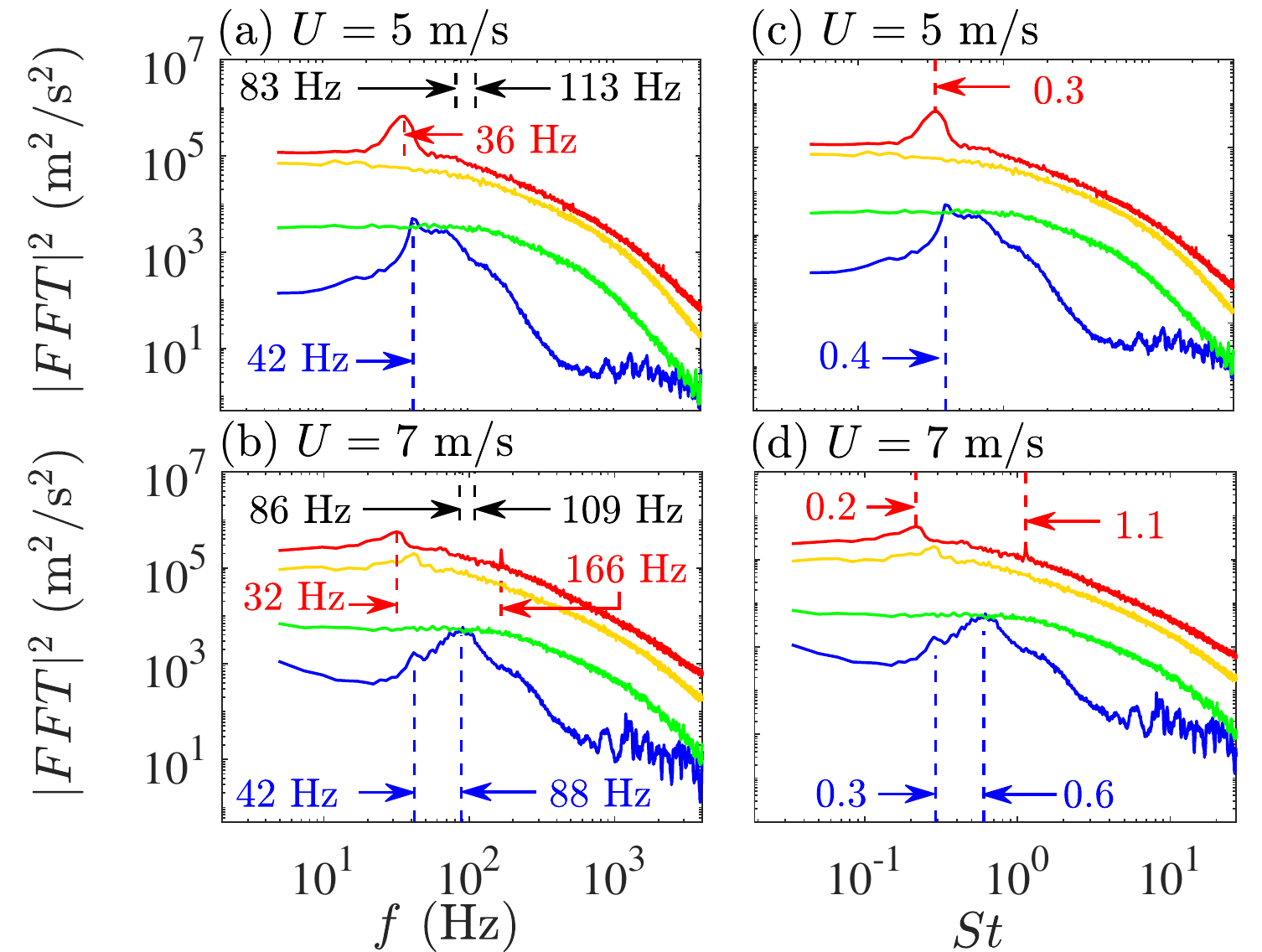}
	\caption{(a) and (b) are the power spectrum densities of the velocity data measured at $x= 0$ and $y=110$~mm corresponding to $U = 5.0$ and 7.0 m/s, respectively. (c and d) are the results in (a and b) presented versus the Strouhal number. The blue, green, yellow, and red colors, pertain to no turbulence generator, one perforated plate, two perforated plates, and the CFSD, respectively.}
	\label{fig:PSD}
\end{figure}

The power spectrum density of the velocity fluctuations at $x= 0$ and $y=110$~mm for test conditions related to $U=5.0$ and 7.0~m/s are presented in Figs.~\ref{fig:PSD}(a) and (b), respectively. The blue, green, yellow, and red colors pertain to no turbulence generator, one perforated plate, two perforated plates, and the CFSD, respectively. For presentation purposes, the PSD was calculated by binning the collected velocity data into 200 segments (4500 data points each), calculating the Fast Fourier Transform (FFT) squared of the velocity fluctuations for each segment, and averaging these. The number of points for the FFT calculation was set to $n_\mathrm{FFT}=2^{12}$. For the results presented in Figs.~\ref{fig:PSD}(a) and (b), the integral of $|FFT|^2$ with respect to frequency normalized by $f_\mathrm{s}\times n_{\mathrm{FFT}}$ equals the variance of the velocity fluctuations. For presentation purposes, the FFT results were truncated at $f = 4000$~Hz (instead of $f = f_\mathrm{s}/2 = 5000$~Hz) since most tested conditions featured noise for $f > 4000$~Hz. The power spectrum density of $u$ for such frequency range is about 4 orders of magnitude smaller than the corresponding maximum $|FFT|^2$. Thus, filtering the velocity data for $f<4000$~Hz did not influence the velocity statistics in the present study and low-pass filtering the velocity data was not necessary and was not performed.

For comparison purposes, the horizontal axes of Figs.~\ref{fig:PSD}(a and b) were normalized and the results are presented in Figs.~\ref{fig:PSD}(c and d). In these figures, the Strouhal number ($St$) equals $fD/U$, with $f$ and $D=47.7$~mm being the frequency and the exit diameter of the turbulence generation section, respectively. In Fig.~\ref{fig:PSD}(c and d), the blue, green, yellow, and red colors pertain to the test conditions with no turbulence generator, one perforated plate, two perforated plates, and the CFSD, respectively. The results show that the PSDs of the velocity data corresponding to one and two perforated plates do not feature a dominant frequency; however, those related to no turbulence generator and the CFSD may feature dominant frequencies. Specifically, for $U = 5.0$~m/s, the PSD of the velocity fluctuations corresponding to no turbulence generator features a dominant peak at 42~Hz, which pertains to the Strouhal number of 0.4, see Fig.~\ref{fig:PSD}(c). At $U = 7.0$~m/s, the free jet features two dominant peaks at 42 and 88~Hz, which correspond to $St=0.3$ and 0.6, respectively, as shown in Fig.~\ref{fig:PSD}(d). The dominant Strouhal numbers of 0.3, 0.4, and 0.6 have been previously observed for free jets and are shown to relate to the vortex shedding from the jet shear layers, see for example, Birbaud \textit{et al.}~\cite{birbaud2007dynamics} and Gutmark~and~Ho~\cite{gutmark1983preferred}.

The PSD of the velocity data corresponding to the Compact Flow Stirring Device features a dominant peak at 36~Hz ($St = 0.3$) for $U = 5.0$~m/s and two dominant peaks at 32~Hz ($St = 0.2$) and 166~Hz ($St = 1.1$) for $U = 7.0$~m/s. For comparison purposes, the CFSD blades rotation frequencies (discussed in Section~\ref{section:tools}) are shown by the black dashed lines in Figs.~\ref{fig:PSD}(a and b). It can be seen that the PSD of the velocity data does not feature a dominant peak at the blades rotation frequencies. It is concluded that, while it may be likely that the instability induced by the CFSD blades rotation is present immediately downstream of the blades, such instability is not observed at the location of the velocity measurements (110~mm downstream of the blades). In essence, despite the Compact Flow Stirring Device generates relatively large values of the RMS velocity fluctuations (compared to past studies that used active generators in turbulent premixed combustion research), the signature of the CFSD blades rotation is not present in the decaying region of turbulence.

\begin{figure}[!h]
	\centering
	\includegraphics[width=1\textwidth]{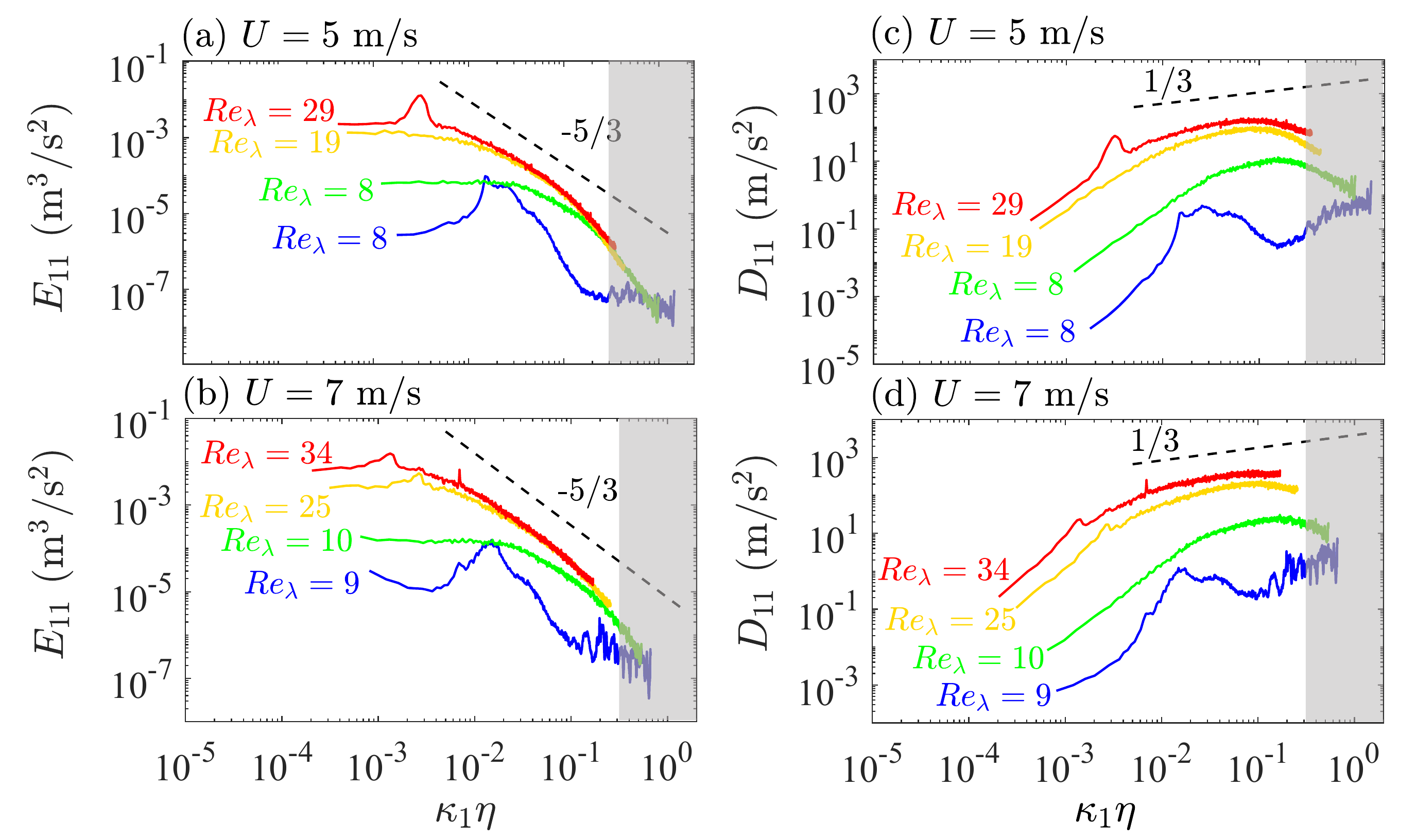}
	\caption{(a) and (b) are the normalized 1D turbulent kinetic energy spectra measured at $x= 0$ and $y=110$~mm for $U = 5.0$ and 7.0~m/s, respectively. (c) and (d) are the normalized 1D dissipation spectra for $U = 5.0$ and 7.0~m/s, respectively. The blue, green, yellow, and red colors pertain to no turbulence generator, one perforated plate, two perforated plates, and the CFSD, respectively.}
	\label{fig:PSDmodified}
\end{figure}

The results presented in Fig.~\ref{fig:PSD} were used to study the normalized one dimensional turbulent kinetic energy spectrum ($E_{11}$) as well as the normalized one dimensional dissipation spectrum ($D_{11}$) of turbulence generated by the CFSD. $E_{11}$ and $D_{11}$ were calculated using~\cite{mcmanus2020simultaneous}

\begin{subequations}
	\begin{equation}
		E_{11} = \frac{\mathrm{PSD}(u')}{n_\mathrm{fft}\kappa_\mathrm{s}},
	\end{equation}
	\begin{equation}
		D_{11} = 2\kappa_1^2 E_{11},
	\end{equation}
\end{subequations}
where $\kappa_1$ is the one dimensional wave number given by $\kappa_1 = 2\pi f/U$, and $\kappa_\mathrm{s}$ is the one dimensional wave number evaluated at $f = f_\mathrm{s}$. The worst resolution of our HWA measurements can be estimated using the active length of the sensor (which is 1.25~mm), corresponds to the test condition of CFSD7, and is given by $\kappa_1\eta = (2\pi/1.25)\times 0.05 \approx 0.3$. Thus, the regions corresponding to $\kappa_1\eta \gtrsim 0.3$ are grayed out in Fig.~\ref{fig:PSDmodified}. The results presented in Figs.~\ref{fig:PSDmodified}(a and b) show that, the CFSD features the -5/3 power-law decay for $10^{-2} \lesssim \kappa_1\eta \lesssim 10^{-1}$. Similar behavior is also observed for one and two perforated plates and for both tested mean bulk flow velocities. The -5/3 power-law decay of the 1D turbulent kinetic energy corresponds to the slope of $2-5/3 = 1/3$ for $10^{-2} \lesssim \kappa_1\eta \lesssim 10^{-1}$ of the 1D dissipation spectra shown in Figs.~\ref{fig:PSDmodified}(c and d). Compared to one and two perforated plates as well as the CFSD related test conditions, the -5/3 power-law decay is absent for 0P5 and 0P7 test conditions since the turbulence does not decay in the location of measurements for no turbulence generator test conditions (as discussed in Subsection~\ref{section:decay}). Similar observations are reported for near field of a free jet, see for example~Gutmark and Ho~\cite{gutmark1983preferred} and Sadeghi and Pollard~\cite{sadeghi2012effects}.

The Reynolds number estimated based on the Taylor length scale for all tested conditions are shown in Fig.~\ref{fig:PSDmodified}. At relatively small 1D wave numbers ($\kappa_1 \eta \sim 10^{-3}$), increasing $Re_\lambda$ increases both $E_{11}$ and $D_{11}$. For relatively large 1D normalized wave numbers ($\kappa_1\eta \sim 10^{-1}$), $E_{11}$ collapses as $Re_\lambda$ increases; however, $D_{11}$ increases. In fact, for the rest of the examined vertical distances ($y>110$~mm), similar observation was made. The relatively large values of $D_{11}$ for the CFSD (compared to the rest of the tested turbulence generators) is consistent with the results presented in Fig.~\ref{fig:dissipationrate}. The variations of $E_{11}$ and $D_{11}$ versus $\kappa_1\eta$ presented in Fig.~\ref{fig:PSDmodified} have two implications for the corresponding turbulent premixed flames. First, the relatively large values of the RMS velocity fluctuations associated with the CFSD are attributed to relatively large scale eddies. Small scale eddies appear to feature turbulent kinetic energies similar to one and two perforated plates. Second, the turbulent kinetic energy is dissipated faster by the eddies that are generated using the CFSD compared to those of one and two perforated plates. This means that the CFSD produces relatively large scale eddies with large turbulent kinetic energy, which is dissipated faster compared to those of the perforated plates.

\section{Concluding remarks}
\label{section:conc}
A new active turbulence generator, referred to as the Compact Flow Stirring Device (CFSD), was developed and characterized in the present study. Hot-wire anemometry and high-speed imaging were performed to study the turbulent flow characteristics and the CFSD dynamics, respectively. For comparison purposes, experiments were performed for the CFSD as well as three other turbulence generation mechanisms (a free jet, one perforated plate, and two perforated plates). Two mean bulk flow velocities of 5.0 and 7.0~m/s were examined.

It was obtained that the CFSD can generate centerline RMS velocity fluctuations up to 1.8~m/s, which is about 7 folds larger than that reported in the literature for active turbulence generators used to produce turbulent premixed flames. Compared to past studies relevant to active turbulence generators, the CFSD produced turbulence within a relatively small space which facilitates implementation of this device for engineering applications. For two tested gaseous fuels (methane and 40\% by volume hydrogen-enriched methane), it was shown that not only stable turbulent premixed flames can be formed by the CFSD, but also the near flash-back equivalence ratio can be improved using the newly developed device.

It was shown that, in the region of investigation, though the free jet turbulence was developing, one perforated plate, two perforated plates, and the CFSD featured decaying turbulence. For these turbulence generators, the integral, Taylor, and Kolmogorov length scales increased along the vertical axis, and the flow was nearly homogeneous. The power-law exponent of the turbulence decay was calculated for one perforated plate, two perforated plates, and the CFSD; and, this parameter was equal to $-1.0$, $-1.8$, and $-1.0$ as well as $-1.0$, $-2.3$, and $-1.3$ for mean bulk flow velocities of 5.0 and 7.0~m/s, respectively. Assuming isotropy, the normalized energy dissipation rate ($C_\epsilon$) and the dissipation rate ($\epsilon$) were estimated. It was obtained that, although the normalized energy dissipation rate is relatively small for the CFSD (compared to those of other tested turbulence generators), the CFSD features the largest values of $\epsilon$.

The Fast Fourier Transform was used to study the spectral content of the velocity fluctuations. It was shown that, while the power spectrum densities of the velocity data corresponding to one and two perforated plates do not feature a dominant frequency, those related to a free jet and the CFSD features dominant frequencies related to the Strouhal numbers of 0.2, 0.3, 0.6, and 1.1, depending on the tested mean bulk flow velocity. For the test conditions related to the CFSD, none of these Strouhal numbers relate to the CFSD blades rotation frequencies. The power spectrum densities of the velocity data were used to calculate the one dimensional turbulent kinetic energy and dissipation spectra. It was shown that, for all tested conditions, except the free jet, the energy and the dissipation spectra follow the -5/3 and 1/3 power-laws for normalized one dimensional wave number ranging from about 0.01 to 0.1. Agreeing with the literature, increasing the Taylor based length scale Reynolds number increased the energy and dissipation rate of the large eddies. For small scale eddies, while the turbulent kinetic energy spectra collapsed, the dissipation spectra increased moving from no turbulence generator to one perforated plate, two perforated plates, and the CFSD. 

Although the developed Compact Flow Stirring Device offered features that removed some limitations for application of active turbulence generators in turbulent premixed combustion research and potential industrial applications, future work needs to be completed to improve the design and the characteristics of the CFSD. It is acknowledged that the robust rotation of the CFSD blades can be impacted at mean bulk flow velocities other than those tested here. This will be addressed in future designs aiming to increase/modulate both the RMS velocity fluctuations and the integral length scales independently.

\section*{Acknowledgments}
The authors are grateful for the financial support from the Natural Sciences Engineering Research Council (NSERC) Canada, Fortis BC, as well as the Canada Foundation for Innovation.

\section*{Appendix A: Control system for the compact flow-stirring device}

\begin{figure}[!h]
	\centering
	\includegraphics[width=0.75\textwidth]{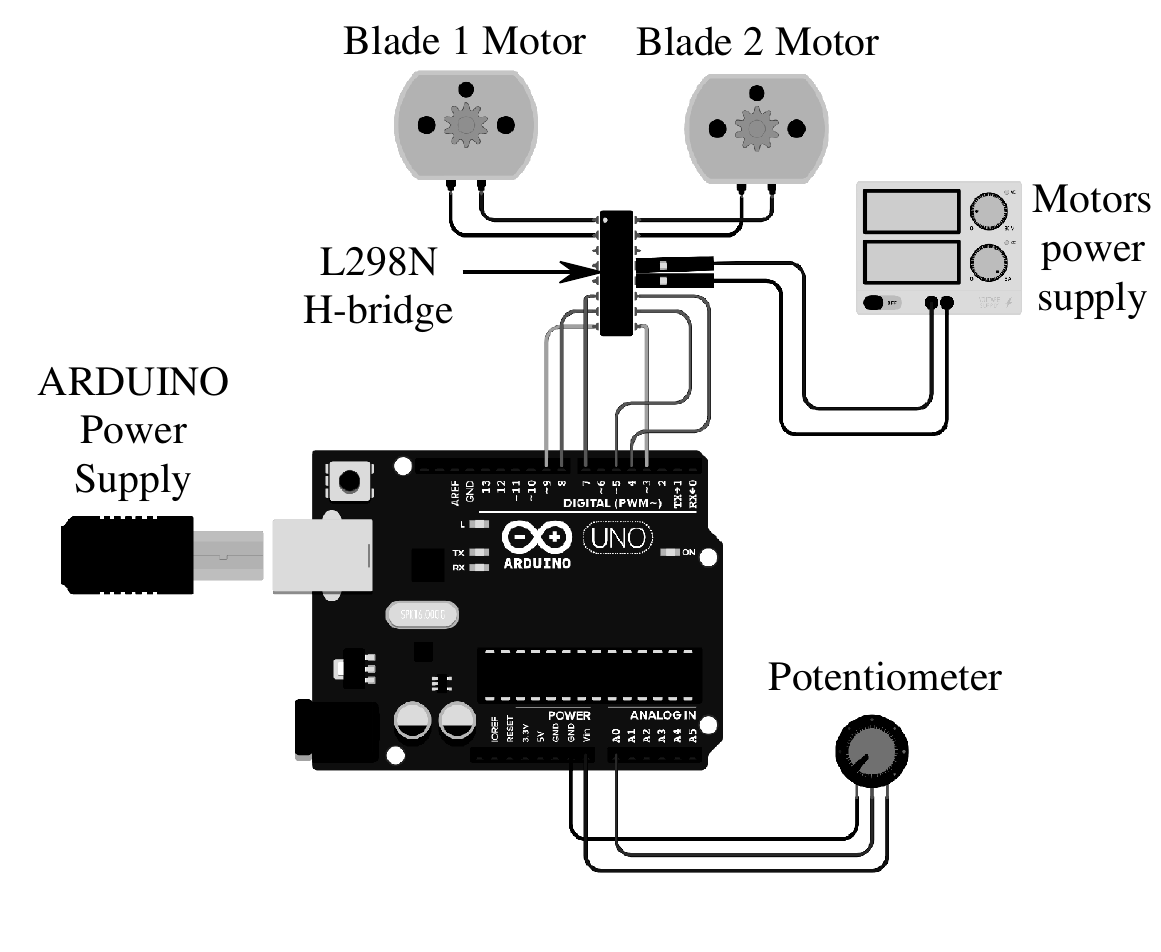}
	\caption{The control equipment used for operation of the CFSD.}
	\label{fig:schem}
\end{figure}

A potentiometer, a motor speed controller (L298N H-Bridge), an ARDUINO UNO board, a power supply, and a routine developed in the ARDUINO UNO language were used to actuate the CFSD motors. The schematic of the control system is presented in Fig.~\ref{fig:schem}. For each test condition that involved operation of the CFSD, the ARDUINO board was powered by an external power supply first. Then, using a potentiometer, the H-bridge was activated and controlled to run two motors that rotated the CFSD blades.

%Line 28 and 33 in the routine below were implemented for safe operation of the motors. If, for unforeseen reasons, such as sudden increase of the mean bulk flow velocity, the rotation speeds of the motors reduced to 5/255 of the set value, the motors stopped actuation. Line 32, allowed for implementing a 500~ms time delay for the motors speeds to reach to their nominal value. This prohibited significant current supply to the motors, which could cause damage.

\section*{Appendix B: Hot-wire calibration and uncertainty of the velocity measurements}

\begin{figure}[!h]
	\centering
	\includegraphics[width=0.8\textwidth]{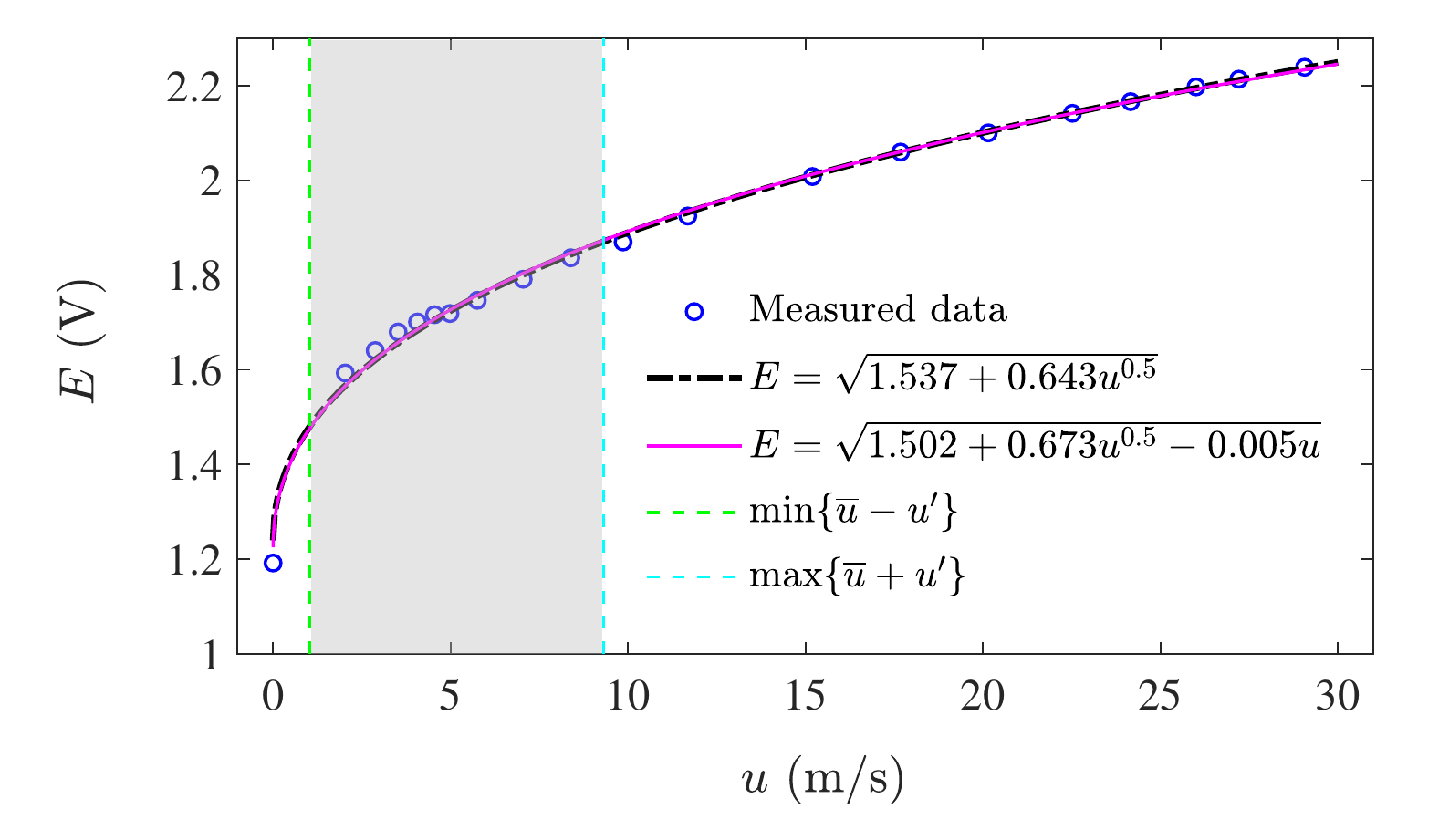}
	\caption{The blue circular data points present the variation of the mean measured voltage versus mean streamwise velocity obtained in calibration of the hot-wire system. The dotted-dashed black and solid purple curves are fits recommended in Bruun \textit{et al.}~\cite{bruun1988velocity}. The region with vertical dashed borders highlights approximate range of the collected velocity data in the present study.}
	\label{fig:calCurves}
\end{figure}

Model 9054H0101 of Dantec was used for calibration of the wire. This model features a mechanical lever for adjusting the flow rate of a free jet, whose centerline velocity is related to the air pressure drop inside the model. The hot-wire measurements were completed during three days, and the calibration was performed on a daily basis before and after the experiments. For each calibration, 20 pressure drops (corresponding to 20 positions of the mechanical lever) were measured using HHP8252 model from Omega, and the corresponding voltage from the mini-CTA was acquired. Then, the measured pressure drops were provided as input to a calibration spreadsheet from the producer, and the centerline jet velocity was generated by the spreadsheet. Following this procedure, 20 voltage data relating to 20 jet centerline velocities were acquired. A sample variation of the measured mean voltage versus the mean velocity is presented in Fig.~\ref{fig:calCurves}, using the open blue circular data symbol. The calibration offset (mean voltage at zero velocity) was measured and equals 1.19~V. Following the recommendations of Bruun \textit{et al.}~\cite{bruun1988velocity}, two curves with the mathematical formulations of 
\begin{subequations}
	\label{Eq:curvefit}
	\begin{equation}
	E = \sqrt{a_0+a_1u^{0.5}},		
	\end{equation}
	\begin{equation}
	E = \sqrt{b_0+b_1u^{0.5}+b_2u}
	\end{equation}
\end{subequations}
were fit to the calibration data using the method of least squares. These fits and the corresponding values of $a_{0}$ and $a_{1}$ as well as $b_{0}-b_{2}$ are presented in Fig.~\ref{fig:calCurves}. The results in the figure show that both formulations accurately predict the relation between $E$ and $u$. In the present study, Eq.~(\ref{Eq:curvefit}a) was used to convert the measured voltages to the velocity data. For all tested conditions, the smallest value of $\overline{U}-u'$ and largest value of $\overline{U}+u'$ were obtained and overlaid on Fig.~\ref{fig:calCurves} using the dashed lines. This range of variation highlights the approximate range of the collected velocity data, and as can be seen, this range is well represented with a large number of calibration data points (9 out of 20).

The overall uncertainty of the hot-wire measurements depends on the bias ($Bi$) and the precision ($S$) errors as discussed in Moffat~\cite{moffat1988describing}. The total uncertainty with 95\% confidence level (referred to as $\delta$ here) is given by~\cite{moffat1988describing} $\delta = \sqrt{Bi^2+(2S)^2}$. The precision error is directly and inversely proportional to the RMS of the velocity fluctuations and the square root of the number of acquired data, respectively. Given the relatively large number of velocity data collected for each test condition (900000 data points), the precision error is negligible in the present study. Thus, $\delta \approx Bi$. The bias error is given by~\cite{moffat1988describing} $Bi =\sqrt{Bi_\mathrm{CAL}^2+Bi_\mathrm{PROBE}^2}$, with $Bi_\mathrm{CAL}$ and $Bi_\mathrm{PROBE}$ being the bias error due to calibration and hot-wire probe positioning, respectively. The bias error due to the calibration is mainly associated with the temperature drift during the experiments and the difference between the measured calibration data and the prediction of the calibration curve, which is obtained from Eq.~(\ref{Eq:curvefit}a) in this study. The calibration curves before and after the experiments nearly collapsed, the bias error due to temperature drift was negligible, and calibration correction such as that in Hultmark and Smits~\cite{hultmark2010temperature} was not required. The bias error due to the difference between the measured calibration data and that from Eq.~(\ref{Eq:curvefit}a) was about 4\% ($Bi_\mathrm{CAL}=4\%$). The positioning of the probe accuracy is mostly influenced by the sCMOS camera pixel resolution (40~$\mu$m). The bias error due to inaccurate positioning of the probe is mostly pronounced close to the shear layers, where the velocity features large gradients. Although utmost care was taken to ensure accurate positioning of the probe, it was estimated that $\pm$40~$\mu$m inaccuracy in positioning of the probe can lead to a maximum of 2\% error in the velocity data, i.e. $Bi_\mathrm{PROBE} = 2\%$. Thus, the overall error in estimation of the velocity data with 95\% confidence level is $\delta \approx Bi = \sqrt{Bi_\mathrm{CAL}^2+Bi_\mathrm{PROBE}^2} = \sqrt{4\%^2+2\%^2} \approx 5\%$.

%\newpage

%\begin{lstlisting}[style=myArduino]
%unsigned long time;
%int ENA = 9;
%int ENB = 3;
%int node1 = 8;
%int node2 = 7;
%int node3 = 5;
%int node4 = 4;
%int speedControl1 = A0;
%int motorSpeed = 0;
%void setup() {
	%	pinMode(ENA, OUTPUT);
	%	pinMode(ENB, OUTPUT);
	%	pinMode(node1, OUTPUT);
	%	pinMode(node3, OUTPUT);
	%	pinMode(node2, OUTPUT);
	%	pinMode(node4, OUTPUT);
	%	Serial.begin(9600);
	%}
%void demoOne() {
	%	digitalWrite(node1, HIGH);
	%	digitalWrite(node2, LOW);
	%	analogWrite(ENA, 10);
	%	digitalWrite(node3, HIGH);
	%	digitalWrite(node4, LOW);
	%	analogWrite(ENB, 10);
	%	motorSpeed = analogRead(speedControl1);
	%	motorSpeed = map(motorSpeed, 0, 1023, 255, 0);
	%	if (motorSpeed < 5){
		%		motorSpeed == 0;
		%	}
	%	analogWrite(ENA, motorSpeed);
	%	analogWrite(ENB, motorSpeed);
	%	delay(500);
	%}
%void loop() {
	%	demoOne();
	%}
%\end{lstlisting}


%\bibliographystyle{abbrv}
%\nocite{*}

\section*{Appendix C: Influences of the initial time and time window size on the integral length scale}

\begin{figure}[!h]
	\centering
	\includegraphics[width=1\textwidth]{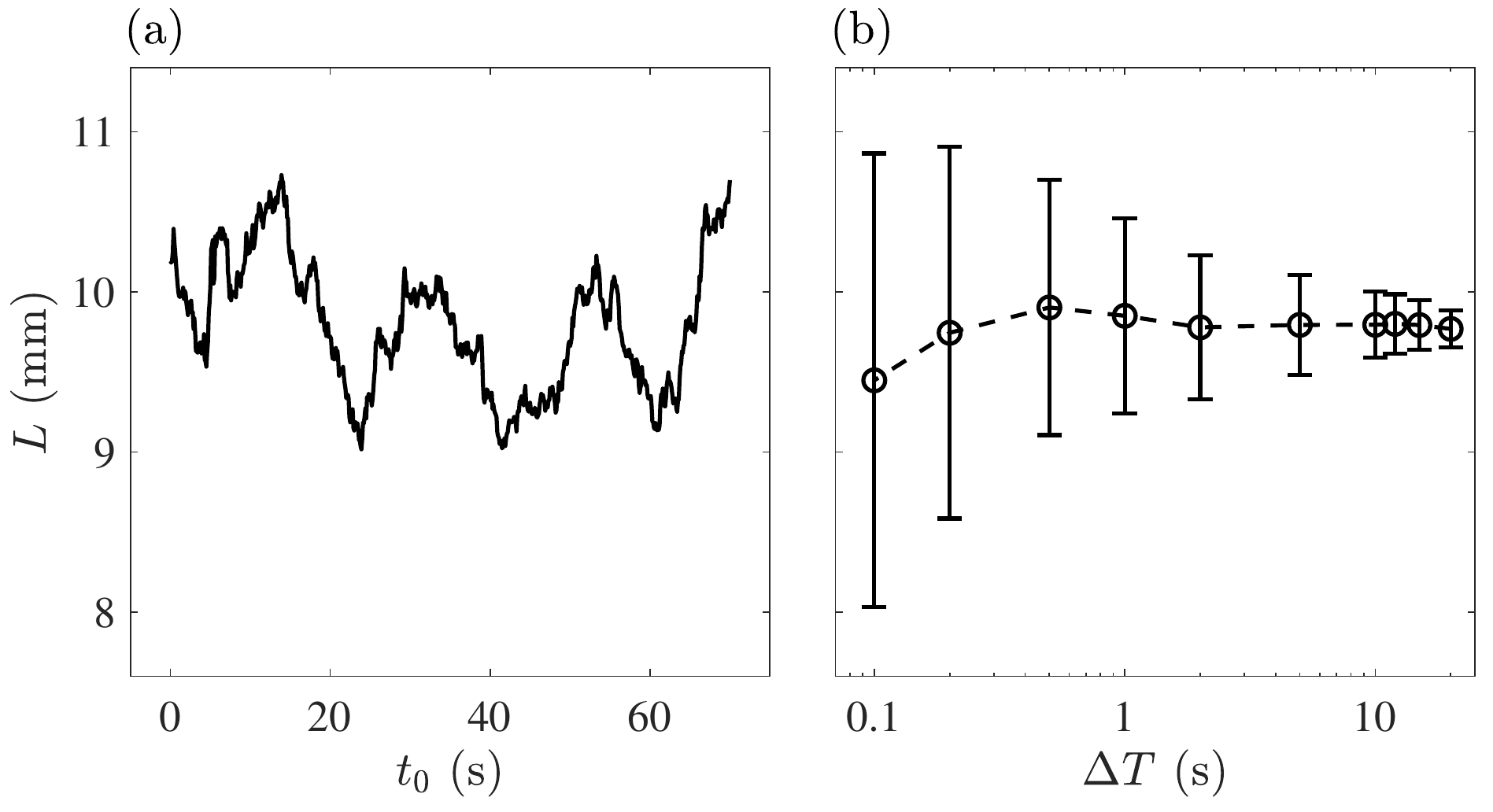}
	\caption{(a) and (b) present the variations of $L$ with $t_0$ and $\Delta T$, respectively. The results pertain to the test condition of CFSD7 and the integral length scales are evaluated at $(x= 0,y = 110~\mathrm{mm})$.}
	\label{fig:WS}
\end{figure}

Figures~\ref{fig:WS}(a) and (b) present the variations of the integral length scale versus $t_0$ and $\Delta T$, respectively. The results pertain to the test condition of CFSD7 and are evaluated at $(x= 0,y = 110~\mathrm{mm})$. This test condition and the spatial location are selected as the corresponding integral length scale featured relatively large sensitivity to $t_0$ and $\Delta T$. For the results presented in Fig.~\ref{fig:WS}(a), $\Delta T = 10$~s. As can be seen, changing $t_0$ changes $L$ from about 9 to 10.7~mm. For the fixed value of $t_0 = 0$~s, the results in Fig.~\ref{fig:WS}(b) show that changing $\Delta T$ from about 0.1 to 20~s does not change the integral length scale appreciably. In fact, this is the case for all tested conditions. In this study, the root-mean-square of $L$ fluctuations estimated for several values of $t_0$ was used to quantify the variability of the integral length scale and equals the size of the error bars in Fig.~\ref{fig:WS}(b).

\bibliography{sn-bibliographyLS2}% Produces the bibliography via BibTeX.
	
\end{document}